\documentstyle[12pt]{article}
\def\pw#1{^{#1}}

\def\cN{{\cal{N}}}
\def\cO{{\cal{O}}}
\def\cJ{{\cal{J}}}
\def\cI{{\cal{I}}}
\def\cR{{\cal{R}}}
\def\cRone{{\cal{R}}_1}
\def\cRtwo{{\cal{R}}_2}
\def\cRi{\cR_{\mbox{\tiny indep.}}}
\def\cRc{\cR_{\mbox{\tiny coher.}}}
\def\Vo{\vec{v}_1}
\def\Vt{\vec{v}_2}
\def\Tcr{\Theta_{\mbox{\tiny crit}}}
\def\Vz{\vec{v}_0}
\def\Vn{\vec{n}}
\def\cT{{\cal{T}}}
\def\bv{\bar{v}}
\def\mt{m_t}
\def\t12{\theta_{12}}

\def\baeq{\begin{appeq}}     \def\eaeq{\end{appeq}}
\def\baeeq{\begin{appeeq}}   \def\eaeeq{\end{appeeq}}
\newenvironment{appeq}{\beq}{\eeq}
\newenvironment{appeeq}{\beeq}{\eeeq}

\renewcommand{\theequation}{\thesection.\arabic{equation}}

\newcounter{hran}
\def\bmini{\setcounter{hran}{\value{equation}}
\refstepcounter{hran}\setcounter{equation}{0}
\renewcommand{\theequation}
{\thesection.\thehran\alph{equation}}\begin{eqnarray}}

\def\bminiG#1{\setcounter{hran}{\value{equation}}
\refstepcounter{hran}\setcounter{equation}{-1}
\renewcommand{\theequation}{\thesection.\thehran\alph{equation}}
\refstepcounter{equation}\label{#1}\begin{eqnarray}}

\def\emini{\end{eqnarray}\relax\setcounter{equation}
{\value{hran}}\renewcommand{\theequation}{\thesection.\arabic{equation}}}

\def\la{\mathrel{\mathpalette\fun <}}
\def\ga{\mathrel{\mathpalette\fun >}}
\def\fun#1#2{\lower3.6pt\vbox{\baselineskip0pt\lineskip.9pt
  \ialign{$\mathsurround=0pt#1\hfil##\hfil$\crcr#2\crcr\sim\crcr}}}

\def\ie{\hbox{\it i.e.}{ }}      
\def\eg{\hbox{\it e.g.}{ }}      \def\cf{\hbox{\it cf.}{ }}
\def\eV{{\rm e\kern-0.12em V}}            
 \def\GeV{{\rm G}\eV} 
\def\half{{\textstyle {1\over2}}}
  
\def\fm{\,{\rm fm}}
\def \al {\relax\ifmmode{\alpha}\else{$\alpha${ }}\fi}
    
    \def\Re{\mathop{\rm Re}}
\def\partder#1{{\partial   \over\partial #1}}
\def\abs#1{\left| #1\right|}
\def\lrang#1{\left\langle #1 \right\rangle}

\def\ben{\begin{enumerate}}  \def\een{\end{enumerate}}
\def\bit{\begin{itemize}}    \def\eit{\end{itemize}}
\def\beq{\begin{equation}}   \def\eeq{\end{equation}}
\def\beeq{\begin{eqnarray}}  \def\eeeq{\end{eqnarray}}
\def\bq{\begin{quote}}       \def\eq{\end{quote}}

\def\kp{\relax\ifmmode{k_\perp}\else{$k_\perp${ }}\fi}
\def\kps{\relax\ifmmode{k_\perp\pw2}\else{$k_\perp\pw2${ }}\fi}
\def \as{\relax\ifmmode\alpha_s\else{$\alpha_s${ }}\fi}
\def \pt{\relax\ifmmode{p_t}\else{$p_t${ }}\fi}

\def\ad{\relax\ifmmode\gamma\else{$\gamma${ }}\fi}
\def\om{\relax\ifmmode\omega\else{$\omega${ }}\fi}
\def\Th{\relax\ifmmode\Theta\else{$\Theta${ }}\fi}
\def\si{\relax\ifmmode\sigma\else{$\sigma${ }}\fi}

\def\antiq{\relax\ifmmode{\overline{q}}\else{$\overline{q}${ }}\fi}

\def\ee{\relax\ifmmode{e\pw+e\pw-}\else{${e\pw+e\pw-}${ }}\fi}
\def\qq{\relax\ifmmode{q\overline{q}}\else{$q\overline{q}${ }}\fi}
\def\QQ{\relax\ifmmode{Q\overline{Q}}\else{$Q\overline{Q}${ }}\fi}
\def\qqg{\relax\ifmmode{q\overline{q}g}\else{$q\overline{q}g${ }}\fi}

\def\lapp{{\ \lower 0.6ex \hbox{$\buildrel<\over\sim$}\ }}
\def\gapp{{\ \lower 0.6ex \hbox{$\buildrel>\over\sim$}\ }}
\newcommand{\gapproxeq}{\lower .7ex\hbox{$\;\stackrel{\textstyle >}{\sim}\;$}}
\newcommand{\lapproxeq}{\lower .7ex\hbox{$\;\stackrel{\textstyle <}{\sim}\;$}}
\def\be{\begin{equation}}
\def\ee{\end{equation}}
\def\to{\rightarrow}

\def\as{\alpha_s}

\def\bbar{{\bar{b}}}
\def\bb{{b\bar{b}}}
\def\tbar{{\bar{t}}}
\def\tt{t\bar{t}}
\def\ww{W^+W^-}

\def\ffff{f\bar{f'}f\bar{f'}}
\def\degree{^{\circ}}
\def\fm{{\rm fm}}

\def\GeV{{\rm GeV}}

\def\ee{e^+e^-}

\def\gt{\Gamma_t}

\newskip\humongous \humongous=0pt plus 1000pt minus 1000pt
\def\caja{\mathsurround=0pt}
\def\eqalign#1{\,\vcenter{\openup1\jot
\caja   \ialign{\strut \hfil$\displaystyle{##}$&$
\displaystyle{{}##}$\hfil\crcr#1\crcr}}\,}

\newif\ifdtup

\def\eqal2#1{\,\vcenter{\openup1\jot
\caja   \ialign{\strut \hfil$\displaystyle{##}$&\hfil$
\displaystyle{{}##}$\hfil &$
\displaystyle{{}##}$\hfil\crcr#1\crcr}}\,}

\def\tdkbis{1}
\def\kos{2}
\def\cdf{3}
\def\kuhn{4}
\def\bigi{5}
\def\orrros{6}
\def\jikia{7}
\def\muta{8}
\def\altarelli{9}
\def\sommerfeld{10}
\def\sakharov{11}
\def\gammaw{12}
\def\wwpaper{13}
\def\book{14}
\def\webber{15}
\def\velt{16}
\def\ghadir{17}
\def\lund{18}
\def\drag{19}
\def\oldqed{20}
\def\study{21}
\def\fujii{22}

\begin{document}
\begin{titlepage}
\vspace*{-1cm}
\begin{flushright}
 DTP/92/88   \\
UCD-92-29 \\
LU-TP-92-33 \\
     December 1992 \\
\end{flushright}
\vskip 1.cm
\begin{center}
{\Large\bf Properties of  Soft  Radiation Near $\tt$
and $W^+W^-$ Threshold}
\vskip 1.cm
{\large Yu.L. Dokshitzer}
\vskip .2cm
{\it Department of Theoretical Physics, University of Lund \\
S\"olvegatan 14A, S-22362 Lund, Sweden }\\
\vskip .4cm
{\large V.A. Khoze}
\vskip .2cm
{\it Department of Physics, University of Durham \\
Durham DH1 3LE, England }\\
\vskip .4cm
{\large Lynne H. Orr}
\vskip .2cm
{\it Department of Physics, University of California \\
Davis, CA 95616, USA. } \\
\vskip .4cm
and
\vskip   .4cm
{\large  W.J. Stirling}
\vskip .2cm
{\it Departments of Physics and Mathematical Sciences, University of Durham \\
Durham DH1 3LE, England }\\
\vskip 1cm
\end{center}
\begin{abstract}

We discuss the characteristic interference
features of soft radiation in the threshold
production of heavy unstable particles:
soft gluon radiation in $\ee \to \tt$ and soft photon radiation
in $\ee \to \ww$. We show that the heavy
particle decay width controls the interference between the  emission
off the final state particles.
As a result,
the radiation pattern may provide a way of measuring the decay width of the
heavy particles.

\end{abstract}
\vfill
\end{titlepage}
\newpage
\section{Introduction}

Heavy unstable charged particles can emit radiation both before and after
they decay. The analysis of such radiation is a complex issue, depending
sensitively on the timescale of the emission compared to the lifetime of
the unstable particle [\tdkbis]. In particular, the radiation pattern can
be very
different according to whether the radiation occurs predominantly during the
production stage or after the particle has decayed [\kos].

There are several important examples of such effects which are directly
relevant to present and future high-energy colliders.
As a specific example, consider the production and decay
of a $\tt$ pair in high-energy $\ee$ annihilation.
With a mass of at least 91 GeV [\cdf], the top quark can decay to a real
$W$ boson and a $b$ quark. The width $\gt$ for this decay is quite large ---
so large that the top weak lifetime can be as short as strong interaction
timescales.  The resulting interplay between the strong and weak interactions
of the top quark gives rise to interesting physical effects.
For example, if top is heavier than $\sim \ 100\ \GeV$, then $\gt$ can be
greater than
the typical hadronic scale $\mu \sim 1 \; \fm^{-1}$
and it may decay {\it before}
it has time to hadronize [\kuhn-\orrros].  In particular, $\tt$ resonances may
never be formed.
Here we are interested in the perturbative aspects
 of the strong-weak interplay:
decay versus gluon bremsstrahlung.
Reference [\kos] discussed soft gluon radiation in $\ee \to \tt$ and showed
that
gluons radiated in top production and decay
can interfere, and how much they do depends on the top width.
This means that top production and decay should not be treated separately --
the gluon distribution in top events is not what one might naively
guess.
Furthermore, this width effect might be useful;  the sensitivity of the soft
gluon distribution to $\gt$ suggests a way to measure it [\kos,\jikia].
 The width dependence of the gluon distribution at
high collision energies was studied in Ref. [\kos]; however, it was
 found that the configurations with the most
sensitivity to $\gt$ were also the least likely to occur.

In this paper we consider on soft gluon radiation near the $\tt$ threshold.
The top quarks are produced nearly at rest and essentially do not
radiate.  The width dependence is a result of interference between gluons
radiated in the two decays, which does not play an important role at
higher energies.  Near the production threshold,
the amount of interference between gluons from the $b$ and
$\bbar$ is controlled by the top width, and what matters is
the size of $\gt$ relative to the gluon energy.
Thus we will see that when the top width and the gluon energy are more or
less the same order of magnitude, the radiation pattern is sensitive to $\gt$.

A second process which exhibits similar features is the emission of photon
radiation in the process $\ee\to\ww\to\ffff$. The radiation pattern of a soft
photon of energy $\omega$ is sensitive to the $W$ decay width for $\omega
\sim\Gamma_W$. Here there is the additional complication of radiation
off the initial state as well, but, as we shall see, this can easily be
taken into account.

In principle, therefore, the study of the soft gluonic and photonic
radiation in $\tt$ and $\ww$ production provides a basis for determining
the decay width of the heavy particle. In practice, however, there are many
difficulties. The measurement of the radiation pattern in top quark production
requires the separation and identification of a soft gluon jet (typically
with energy $\omega \sim 5\ \GeV$). In the case of photons radiated in
$\ww$ production, while identification of relatively soft photons might
not pose too many problems, the event rates are low for the
anticipated luminosities of future $\ee$ colliders.
Nevertheless, we believe these issues are worth exploring for several
reasons. First, on a theoretical level there are several features
of the radiation patterns that show interesting
interference  effects which are at first sight counter-intuitive.
Second, the `traditional' methods of measuring the masses and decay widths by
threshold scanning are not without their own problems.
Especially for  $\tt$ production
at high-energy $\ee$ linear colliders, the structure of the
threshold is smeared by beam-induced effects,
 intrinsic energy spread
etc.
On the theoretical level, the measurement of $\gt$ from the shape of the cross
section as a function of beam energy near threshold is a delicate
issue.
For $\ee\to\ww$, the threshold scan strategy requires a detailed
calculation of higher order electroweak effects including width effects
and initial state
radiation. To our knowledge, a comprehensive  one-loop calculation including
the
width effects has not yet been completed. An attempt to incorporate
$\Gamma_W$ into the tree-level formulae has been made in Ref. [\muta]
(see also Ref. [\altarelli]). These results however are seriously
affected by
initial state radiation and other effects like the
final-state Coulomb attraction near the $W^+W^-$ threshold
[\sommerfeld, \sakharov].
As a general comment, one might argue that the $W$ width is already
known to fairly high precision ($\Gamma_W = 2.15 \pm 0.11\ \GeV$ [\gammaw])
from {\it indirect} measurements using $W$ production cross sections
in $p\bar{p}$ colliders. Nevertheless, we believe it is important
to obtain a {\it direct} measurement of this important Standard Model
parameter as well.

In this paper, therefore, we will focus mainly on the theoretical features
of soft radiation in $\tt$ and $\ww$ production. Some illustrative numerical
results for $\tt$ production are presented; a more complete
numerical treatment of the $\ww$ case, particularly with reference to LEP200,
is deferred to another paper [\wwpaper]. We believe that our
conclusions will show that a more detailed experimental study (event rates,
detector capabilities, etc.)  is certainly warranted.
The remainder of the paper is organized as follows.  In Sections 2 and 3
 we discuss
the radiation pattern, for gluons in $\tt$ production
and photons in $\ww$ production respectively,  near threshold in detail.
  In Section 4 we present numerical results   for
top production and discuss prospects for measuring
$\gt$.
We conclude in Section 5.  Appendices contain a semi-classical
derivation of the radiation pattern and further details of the calculation of
the distributions.

\section{Soft radiation pattern}

We are interested in emission of a gluon in the process
$\ee \to \tt \to W W \bb$ and of a photon in the process
$\ee \to \ww \to \ffff$. Although the analysis of the final
state radiation in both processes is very similar, the latter process
is complicated by the additional contributions from initial state
radiation.  To begin with, therefore, we discuss the $\tt$ case, and
extend the analysis to $\ww$ production in the next Section.

The general result for soft gluon radiation in $\ee \to\tt \to WW \bb$
was presented in reference [\kos] (see also [\jikia]).
Here we focus on the particular case
of radiation close to the $\tt$ threshold. There are two advantages
in this. First, the production cross section is largest just above threshold.
Second, near  threshold the top quarks are almost at rest and only
the $b$ and $\bbar$ radiate.
While it is not obvious that the top quark width should enter at all
if only the b-quarks radiate, we can understand the its role
as follows.
Consider two cases of gluon radiation from a $\bb$ pair.
If the quarks could radiate independently, with no interference, the
gluon distribution would be proportional to
\begin{equation}
\cRi\; =\; \cRone\> +\> \cRtwo\;
=\; \frac{v_1\pw2\sin\pw2\theta_1}{(1-v_1\cos\theta_1)\pw2}\> +\>
\frac{v_2\pw2\sin\pw2\theta_2}{(1-v_2\cos\theta_2)\pw2}\>,
\label{indep}
\end{equation}
where $v_i$ is the velocity of the $b$ ($\bbar$), and
$\theta_i$ is the angle between the $b$ ($\bbar$) and the gluon for $i=1$
($2$).
(Note that in what follows we will make a distinction between $v_1$ and $v_2$
although in practice, for the case of $e^+e^-\to\tt \to W^+W^-\bb$ in the
centre-of-mass frame, we always have $v_1=v_2$.)
In the other extreme, with interference we have coherent emission, and the
gluon distribution looks like
\begin{eqnarray}
\cRc
     &=&  \cRi + 2 \cJ, \nonumber \\
\cJ \> &\equiv & \>
\frac{v_1v_2(\cos\theta_1\cos\theta_2-\cos\theta_{12})}
{(1-v_1\cos\theta_1)(1-v_2\cos\theta_2)}\; ,
\label{coher}
\end{eqnarray}
where $\theta_{12}$ is the angle between the $b$ and the $\bbar$.  This
is just the familiar antenna pattern for emission from a quark-antiquark
pair; the interference is $2 \cJ$.  Note that these patterns can be quite
different and the interference can be constructive or destructive.
In particular, and as we shall discuss in more detail later (and see
Appendix~E),
coherent emission exhibits angular ordering
behavior, {\it i.e.} if we integrate
over the azimuthal angle about the direction of quark
 1, all radiation from quark 1 is suppressed for angles
$\theta_1$ such that [\book,\webber]
\begin{equation}
\label{angord}
v_2 \cos\theta_{12} < \cos\theta_1 < \cos\theta_c\simeq v_1 \; .
\end{equation}
The factor $v_1$ on the right-hand-side takes account of the screening
effects due to the mass of quark 1 (dead-cone).

Now recall that in the case of interest the $b$ and $\bbar$ are produced
by the decays of the $t$ and $\tbar$.  If the top lifetime is very
short compared to the characteristic time for emitting a gluon
of energy $\omega$ ({\it i.e.}  $\gt \gg \omega$),
the $b$ and $\bbar$ are produced nearly instantaneously,
and we expect coherent gluon emission, with the gluon distribution
determined by $\cRc$.  If however the top lifetime is very long ({\it i.e.}
$\gt \ll \omega$), the $b$ and $\bbar$ are produced at very
different times and thus will radiate independently; the gluon distribution
is then given by $\cRi$.  The ratio $\gt/\omega$ of the top width
to the gluon energy controls the amount of interference.
The full distribution we show below was derived from standard
Feynman diagram techniques in Ref. [\kos].  But in fact it
can be derived from simple semi-classical wave arguments --
see Appendix~A.

Following reference [\kos], the radiation pattern can be presented as a
probability density, normalized to the lowest order cross section:
\beq
dN \equiv 1/\sigma_0 d\sigma_g =
 \frac{d\omega}{\omega}\frac{d\Omega}{4\pi}\>\frac{C_F\as}{\pi}\>\cN\>.
\label{dN}
\eeq
where $C_F = 4/3$ is the QCD colour factor.
Near  threshold,  we have
\begin{eqnarray}
\cN &\equiv&
(1-\chi (\omega ))\cdot\cRi \>+\>\chi (\omega )\cdot \cRc\> \nonumber \\
& = & \cRi + 2 \chi (\omega ) \cdot \cJ \; ,
\label{splitJa}
\end{eqnarray}
where $\cRi$ and $\cRc$ are given in eqns.\ref{indep} and \ref{coher} (with
$v_1 = v_2 = v_b$) and we have introduced the profile function
\beq
 \chi (\omega)  = { \Gamma^2 \over \Gamma^2 + \omega^2}  \; ,
\eeq

The factor $ \chi$, which depends on the decay width
and gluon energy, determines the amount of
interference, and as stated above we have in the limits of large and
small width,
\bmini
\label{ind}
\cN=\cRi && \mbox{for}\quad \omega\gg\Gamma\>,\\
\label{coh}
\cN=\cRc && \mbox{for}\quad \omega\ll\Gamma\>.
\emini
Evidently, we have
maximal sensitivity to $\gt$ for $\omega\sim \gt$, and this provides
a possible basis for measuring the width.
For gluon emission we must in addition impose
 $\omega > \mu \sim 1\ {\rm fm}^{-1}$,
in order to remain in the perturbative regime.

Before performing a detailed numerical study of the above to investigate
the actual sensitivity to $\gt$ in different angular configurations,
we make some additional  comments.

\bit
\item[{(i)}]
The numerator of the $\cJ$ term in  (\ref{coher}) can be written as
\beq\label{dnuma}
\cos\theta_1\cos\theta_2-\cos\theta_{12}=-\sin\theta_1\sin\theta_2\cos\phi_{12}
\eeq
with $\phi_{12}$ the relative azimuth between $\vec{p}_1$ and
$\vec{p}_2$ with respect to the gluon direction, $\vec{k}$.
Thus an immediate consequence of the interference is that the
azimuthal symmetry of the radiation about the $b$-quark
directions is destroyed. Note also that
the expression (\ref{dnuma}) vanishes when the direction of the gluon momentum
is chosen to be close to that of one of the quarks, $\theta_1\ll\theta_{12}$
or $\theta_2\ll\theta_{12}$.
Therefore the $\omega$-dependence of the radiation pattern (\ref{splitJa})
can reveal itself
only if the gluon emission angles are not small compared to the opening
angle of the $b$ and $\bar{b}$ .

\item[{(ii)}]
It is fairly straightforward to integrate over the angle of the emitted gluon
and obtain the {\it total} probability for the  radiation of
a gluon of  given energy. (In a sense this is a formal procedure, since
in practice the gluon jet will only be identified when separated
 from other final
state jets and from the beam.) Full details are given in Appendix~B -- here
we present only the result:
\beeq
\frac{dN}{d\omega} \> &=&\> \frac{C_F\alpha_s}{\pi\omega}\>
\left\{ (1-\chi(\omega )) \left[ \frac{1}{v_1}\log\frac{1+v_1}{1-v_1}
+  \frac{1}{v_2}\log\frac{1+v_2}{1-v_2} -4 \right] \right. \nonumber
\\
&&\left.  \> + \chi \left[ \frac{1}{r}\log\frac{1+r}{1-r} -2
 \right] \right\} \> .
\label{intall}
\eeeq
where
\beeq
 r\> & = & \> \frac{\sqrt{\psi\pw2-1}}{\psi}
\nonumber \\
 \psi \>&=&\>
\frac{p_1\cdot p_2}{M_1M_2}
 \>=\>
\frac{(1\!-\!v_1v_2\cos\theta_{12})}{\sqrt{(1\!-\!v_1\pw2)(1\!-\!v_2\pw2)}}\>.
\eeeq
Note the logarithmic collinear singularities which dominate the integrated
distribution in the ultra-relativistic limit $v_i \to 1$.
In fact for $(1-v_i) \ll 1$ and for $\theta_{12}$ values not particularly
close to $0$, we have (Appendix~B)
\beq
\frac{dN}{d\omega} \> \approx\> \frac{C_F\alpha_s}{\pi\omega}\>
\left\{ \ln\frac{2}{1\!-\!v_1} + \ln\frac{2}{1\!-\!v_2} -4
+ 2\chi(\omega) \left[ \>\ln\frac{1-\cos\theta_{12}}{2} + 1
\right] \right\} .
\label{installlim}
\eeq
Note  that the second $\omega$-dependent
term in (\ref{installlim})
{\em enhances}\/  or {\em depletes}\/ the radiation
according to whether
 $\theta_{12}$ is larger or smaller than $\Tcr \approx 75\pw0$ (Appendix~B).
\eit

\section{Soft photon radiation in $\ee\to W\pw+W\pw-$}

Before presenting our numerical results for the $\tt$ case
 we discuss in this section the extension of the above analysis
to soft photon radiation near threshold in the process
$\ee \to \ww \to \ffff$. Once again the radiation pattern includes
contributions from the $WW$ production and decay antennae, together
with interferences between them. As was demonstrated in Ref.~[\velt],
the radiation at the production stage (the $\widehat{WW}$ antenna)
is given by the classical current expression, as for the $\widehat{\tt}$
antenna [\kos], irrespective of the choice of  gauge.
Apart from overall couplings, colour factors and charges, there are
only two main differences. First, there are additional contributions
from initial state radiation. These pose no particular problems in
practice as long as the final state particles, including the
photon, are kept well away from the beam direction. As shown in
reference [\ghadir], when all the final state particles are
exactly transverse to the beam direction, the initial state radiation
simply adds a small, constant ``background term'' to the radiation pattern,
which is of course independent of the decay width of the decaying
$W$'s. We note also that near the $WW$ threshold there are kinematic
constraints which are different for initial and final state radiation.
The former is limited by the maximal kinematically  allowed
energy
\begin{equation}
\omega \> < \>\omega_{max}\pw{\mbox{\tiny IS}}\>=\> M_W v_W\pw2\>\ll\> M_W \>,
\label{omegamax}
\end{equation}
whereas the constraint  on the energy emitted in the course of decay
is much less severe,
\begin{equation}
\omega\><\>\omega_{max}\pw{\mbox{\tiny FS}}\>=\> \frac{M_W\pw2
-m_0\pw2}{2M_W}\sim M_W \> ,
\end{equation}
where $m_0$ is the minimal invariant mass of the $W$ decay products.

Since we are mainly interested in the region of photon energies
$\omega\sim\Gamma$
where the width has an important effect  on the radiation pattern,
we can imagine choosing a kinematic region,
\beq
 \omega\sim\Gamma \ll M_W\,v_W\pw2 \ll M_W\>,
\eeq
where the soft bremsstrahlung approximation can be used for both
initial and final state radiation, without having to worry
about the kinematic restrictions on the photon energy.
A derivation of the radiation pattern for this situation using the
classical picture considered in Appendix~A is presented in Appendix~C.

A second difference arises when we consider extending the analysis
to {\it hadronic} $W$ decays, {\it i.e.} to $W \to q {\bar q}'$.
This is of course the dominant $W$ decay channel -- about $44\%$
of $WW$ pairs decay to a four-jet final state, and only about $5\%$
of the decays have purely leptonic ($e$ or $\mu$) final states.
In order to achieve a measureable event rate, therefore, it will
probably be necessary to demand at least one hadronically decaying~$W$.

For $W\to q\bar{q}'$
both decay products can now radiate, and the pattern of  radiation
is correspondingly more complicated.
Thus the electromagnetic current caused by the leptonic decay
$W\pw+(q) \to e\pw+(p) \nu_e(\bar p)$,
\beq
\label{twocurrL}
j_\ell^\mu =  \frac{p^\mu}{(kp)}  - \frac{q^\mu}{(kq)} \>,
\eeq
becomes, for the hadronic decay,
$W\pw+(q) \to u(p) {\bar d}(\bar p)$ ,
\beq
\label{twocurrH}
j_h^\mu =  Q\,\frac{p^\mu}{(kp)} + (1\!-\!Q)\,\frac{\bar{p}^\mu}{(k\bar{p})}
- \frac{q^\mu}{(kq)} \>,
\eeq
where $Q=2/3$ is the electric charge of the $u$ quark.
The general final state radiation pattern can then be formed from these
currents
in the usual way,
\bminiG{twoforms}
\cN &\propto& (j_1^\mu e_\mu)^2 + (j_2^\mu e_\mu)^2  -2\chi(\omega)\cdot
(j_1^\mu e_\mu) (j_2^\mu e_\mu)  \\
   &=& (1-\chi(\omega))\cdot\left[\> (j_1^\mu e_\mu)^2 + (j_2^\mu
e_\mu)^2   \>\right]
+ \chi(\omega) \cdot [(j_1^\mu-j_2^\mu)\,e_\mu]^2\>.
\emini

Equivalent formulae to those considered above for gluon radiation
in $t$ decay can then be derived. The analysis is simplest if
we assume that  in each $W$ decay the quark and the accompanying
antiquark are {\it anti-parallel}, otherwise additional angles
have to be introduced; this is certainly justified if the $W$'s are
produced at rest. If we make the further assumption that
the velocities of the quark and antiquark in the decay are equal,
and that experimentally quark and antiquark jets cannot be distinguished,
then we obtain the radiation patterns for the two-quark and the
one-quark-one-leptonic decays (see Appendix~D):
\beeq
\label{Nqqeq}
\cN^{(qq)}\ & = &\   \cRone\;
\left[ \frac{(2Q\!-\!1 )^2+ v_1^2\cos^2\theta_1}
{(1 + v_1\cos\theta_1)\pw2}\right]
\> +\> \cRtwo\;
\left[ \frac{(2Q'\!-\!1 )^2+ v_2^2\cos^2\theta_2}
{(1 + v_2\cos\theta_2)\pw2}\right]
 \nonumber \\
& &+\>  2\; \chi(\omega)\cdot \cJ\;
 \left[\frac{ v_1 v_2 \,\cos\theta_1\,\cos\theta_2}
{(1+v_1\cos\theta_1)\, (1+v_2\cos\theta_2) } \right]  ;    \\
\label{Nqleq}
\cN^{(q\ell)}\ & = & \  \cRone\;
\left[ \frac{(2Q\!-\!1 )^2+ v_1^2\cos^2\theta_1}
{(1 + v_1\cos\theta_1)\pw2}\right]
\> + \>  \cRtwo     \nonumber \\
& &+ \> 2\; \chi(\omega)\cdot \cJ
  \left[\frac{ v_1\cos\theta_1 }{1+v_1\cos\theta_1} \right]\; ,
\eeeq
where $Q=Q'=\frac{2}{3}$ and $v_1, v_2$ denote the velocities
of the up-type quark and antiquark respectively for the $(q\bar q)$
case, and the quantities $\cRone$, $\cRtwo$ and $\cJ$ are defined in
(\ref{indep}) and (\ref{coher}).  These expressions are the
analogues of  the $\tt$ result (\ref{splitJa}) derived above.
The differential cross section is obtained in the same
way, with the substitution $C_F \alpha_s \to \alpha$.

To study the $\theta_{12}$ dependence of the {\it total} photon yield
(again ignoring the isolation cuts which will be required in practice)
 one has to evaluate the integrals over the photon radiation angle
of the interference terms in (\ref{Nqqeq},\ref{Nqleq}).
These integrals are finite at $v_1=v_2=1$ and so we can use the
ultra-relativistic approximation  in this case.
As shown in Appendix~D,
the total photon yield takes the form
\beq\label{ftyWW}
\omega\,\frac{dN\pw{(\al\beta)}}{d\omega}
= \frac{\al}{\pi }\; \cdot \; B^{\al\beta}
\; \cdot \; \cT\pw{(\al\beta)}
\left(\theta_{12} \,,\> \frac\omega\Gamma \right)
\eeq
where we denote by $(\al\beta)$ decay channels of the $W\pw+W\pw-$ system
\beq\label{chan}
\eqal2{
(\al\beta) \>=\> &(\ell\ell)\>:\qquad & W\pw+W\pw-\to \ell\nu + \ell\nu\>,\cr
                 &(q\ell)\>:\qquad    & W\pw+W\pw-\to \ell\nu + \qq \>,\cr
                 &(qq)\>:   \qquad    & W\pw+W\pw-\to \qq + \qq \>.
}\eeq
The branching ratios are approximately $B^{\ell\ell} = 4/81$, $B^{q\ell}
 = 24/81$ and $B^{qq} = 36/81$. for $\ell = e, \mu$.
We have
\bminiG{cNall}
\label{cNll}
\cT\pw{(\ell\ell)} &=& \cRi\pw{(\ell)} + \cRi\pw{(\ell)}  +2\chi(\omega)\cdot
\left[\>\ln\frac{1-\cos\theta_{12}}{2} + 1\>\right]\>, \\
\label{cNql}
\cT\pw{(q\ell )} &=& \cRi\pw{(q)} + \cRi\pw{(\ell)}  +2\chi(\omega)\cdot
\left[\>\ln\frac{\sin\theta_{12}}{2} + 1\>\right]\>, \\
\label{cNqq}
\cT\pw{(qq)} &=& \cRi\pw{(q)} +  \cRi\pw{(q)}  +2\chi(\omega)\cdot
\left[\>\ln\frac{\sin\theta_{12}}{2} + 1\>\right] \>,
\emini
with the $\theta_{12}$-independent contributions
\bminiG{cRilq}
\label{cRil}
\cRi\pw{(\ell)}\> & =& \>\cI(v) \> = \>
 \frac1v\ln\frac{1+v}{1-v} -2\approx
\ln\frac2{1-v} -2\>, \\
\label{cRiq}
\cRi\pw{(q)} \>
&=& \> Q\pw2\cI(v) + (1\!-\!Q)\pw2\cI(\bv) - 2Q(1\!-\!Q) \>+\>
\cO\left( {1\!-\!v\pw2, 1\!-\!\bv\pw2}\right) .
\emini
In (\ref{cRil}),  $v$ is the velocity of a charged lepton; $v$ and $\bv$ in
(\ref{cRiq}) are the velocities of the {\em up}-type quark (antiquark) and,
respectively,  {\em down}-type antiquark (quark) originating from
$W\pw+$ ($W\pw-$).
Once again, the $\omega$ dependent terms in (\ref{cNll}) can be either
positive (larger $\theta_{12}$) or negative (smaller $\theta_{12}$).
For the double-leptonic channel the critical angle is the same
as for the
$\tt$ case ($\approx 75\pw0$) whereas with at least one hadronic decay
 channel it follows from (\ref{cNql})
and (\ref{cNqq}) that the corresponding critical angle is approximately
$47\pw0$.

Unfortunately, the
pure ``partonic'' prediction (\ref{cRiq}) for the independent photon radiation
off the quark-antiquark
antenna is too naive in practice, since quark velocities are not
well-defined for light quarks, and the result takes no account of the
hadronization process where integer charge hadrons are formed and indirect
photons from hadron decays appear.

However these complications do not affect
the main physical property of (\ref{cNall}), namely the fact that the
$\theta_{12}$-dependent (and thus the $\Gamma$-dependent) part of the photon
radiation pattern  {\it is} under control.
As long as $W\pw+$ and $W\pw-$ initiated jets evolve independently from one
another,
the extra yield of indirect photons
remains insensitive to the event geometry.

Finally, we note that an exactly analogous study could be performed
for the soft photon radiation pattern  in the process $\ee\to Z^0 Z^0
\to f\bar f f' \bar{f}'$. The four charged particles in the final state
give rise to a rich  interference structure,
as for the four-jet decays of the $WW$ pair discussed above.
The four charged leptonic decays of the $Z^0Z^0$ pair would
provide a particularly clean environment in which to study
this, but in practice the event rates would be prohibitively low.

\section{Numerical results for top}

\subsection{Preliminary remarks}

In this Section we illustrate the behavior of soft radiation near threshold
with
some examples.  We will discuss the gluon distributions in $\tt$ events case in
some detail, and show a single example for photons
in the $\ww$ case.  Our emphasis will be on the influence of the
width $\Gamma$ on the radiation pattern.
As discussed above, one might expect that, because of the top quark's large
width (and hence short lifetime), the $\bb$ pair would radiate coherently,
as if they had been produced directly.
We will see that the
correct soft gluon distributions can differ considerably from those
which arise from that expectation.  We will explore
the sensitivity of the radiation patterns to $\Gamma$
and at the end we will consider briefly
to what extent this sensitivity might be useful for measuring $\Gamma$.

Before presenting our numerical results it is helpful to summarize the
results from Section 2 that are most relevant to what follows.  Recall that
the gluon emission probability density is given by Eq. (\ref{dN}) with
\beq
\label{N}
\cN = \frac{v^2\sin^2\theta_1}{(1-v\cos\theta_1)^2} +
\frac{v^2\sin^2\theta_2}{(1-v\cos\theta_2)^2}
+
2 \chi(\omega )\,
\frac{v^2(\cos\theta_1\cos\theta_2-\cos\theta_{12})}
{(1-v\cos\theta_1)(1-v\cos\theta_2)},
\eeq
and $\chi(\omega )
 \equiv { \Gamma^2 \over \Gamma^2 + \omega^2}$; $0 \leq\chi\leq1$.
The profile function $\chi$ determines the amount of interference between
radiation off the $b$ and $\bbar$:  for $\Gamma=0$, we have $\chi=0$ and
the interference is suppressed, whereas for $\Gamma\gg\omega$, $\chi\to 1$\
and we have the full coherent emission mentioned in the previous paragraph.
Finally, note again that
\begin{itemize}
\item The only dependence on the relative geometry of the $b$ and $\bbar$
appears in the interference term; independent emission has no $\t12$
dependence.
\item In contrast to the high energy case [\kos] where we saw
mostly destructive interference, the interference term here can have either
sign.  We will see explicitly below that, as pointed out earlier,
it tends to be constructive for small
$\t12$ and destructive for large $\t12$.
\item From the form of $\chi$ it is clear that we have maximum sensitivity to
the width when $\Gamma\sim\omega$ ($\chi$ not near 0 or 1),
{\it i.e.,} when the energy of the radiation is comparable to the width.
\end{itemize}
The remainder of Section 4 amounts to an elaboration of these points.

Now, these interference width effects influence the radiation
pattern, as illustrated in the differential distributions we
present below.  The behavior
discussed above would be clearly evident if
the  differential distributions were observable, {\it i.e.} if
we had
access to arbitrary gluon energies and arbitrarily
large numbers of events, and we could observe partons directly.
Of course, in the real world there are jets and limited statistics.
Therefore in what follows we make some concessions to reality by also
considering
integrated distributions.  Furthermore we assume that the gluon will be
\lq detected' as a soft jet, and so we take 5 GeV as the  minimum
observable gluon energy.
However, this is no substitute for realistic simulations, nor is
it meant to be; our results are meant to suggest the kinds of distributions
that would be interesting to study.

The results presented below are more or less independent of the
top mass.  If we treat $\Gamma$ as a parameter, $m_t$ comes into
(\ref{N}) only weakly through the $b$ quark velocity
$v$.  For definiteness, we use $v=0.9944$,
which corresponds to $m_t=140\ \GeV$ for $m_b=5\ \GeV$ and $m_W=80\ \GeV$.
A further generalization is possible if $\omega$ is kept fixed:
then $\Gamma$ and $\omega$ only enter through $\chi$.
Now, what values of $\chi$ are relevant for accessible gluon
energies and interesting
top masses?  For our canonical $\mt=140\ \GeV$, $\Gamma\approx 0.7\ \GeV$
in the Standard Model.  Taking $\omega=5\ \GeV$, we
obtain $\chi=0.02$, which is rather close to zero.  The gluon distribution
is  therefore close to that for {\it independent} emission,
not coherent emission as we might naively guess.
The story is slightly different for
larger $\mt$ because the Standard Model width grows as $\mt^3$.
For example, for
$\mt=200\ \GeV$, $\Gamma$ approaches 3 GeV and with $\omega=5\ \GeV$
we have $\chi\approx0.3$.  So the heavier the top quark, the
larger are the relevant values of $\Gamma$ and  $\chi$.

\subsection{Soft gluon distributions}

\subsubsection{Radiation out of the $\bb$ plane}
$\cN$ takes a particularly simple form for gluons radiated
perpendicular to the plane of the $\bb$ pair.
In this case,
$\theta_1=\theta_2=\pi/2$ and
\beq
\cN\: \propto\: 1-\chi\cos\theta_{12}.
\eeq
The width and energy dependence are given, via $\chi$, by
the extent of the deviation from
constant behavior of the gluon
emission probability as a function of $\t12$.  This is illustrated
in Fig. 1, where we show\footnote{The factor of $\omega$ in the normalization
is chosen so
that the only $\omega$ dependence is in $\chi$, {\it cf.} Eq. (\ref{dN}).}
 $\omega (dN/d\omega d\Omega)$ as a function of
$\theta_{12}$ for $\chi=0,{1\over 3},{2\over 3},\ {\rm and}\ 1$.
Interference gives rise to a dramatic
difference between the independent emission ($\chi=0$) and coherent
emission ($\chi=1$) cases.  Note that, as indicated above,
the distribution for $\chi=0.02$ for our canonical $140\ \GeV$ top
will be very different from the expected coherent emission case,
and a heavier top with a larger width would lie somewhere in the middle.

We can turn the discussion around to ask under what circumstances
we are sensitive to the exact value of $\Gamma$.
As we have emphasized above and as is clear from Fig. 1, we have maximum
sensitivity for $\chi$ not too
close to 0 or 1.  So for $\mt=140\ \GeV$
distributions of 5 GeV gluons are not very sensitive to
a Standard Model $\Gamma$, but
if we could observe 1 GeV gluons or if the width were much larger than the
Standard Model value, $\chi$ would be in a more sensitive range.  For a
heavy top, though, soft gluons are sensitive to the Standard Model width,
which follows
from the value of $\chi$ in the 200 GeV example above.

Now this out-of-plane distribution has clear, simple properties, but it
is a differential distribution -- a snapshot.  Let us consider
to what extent the characteristics of Fig. 1 are retained in
integrated distributions.  In Figure 2 we show the distribution
of gluon radiation out of the $\bb$ plane, now integrated over $\pm\pi/8$
in each angular direction and over gluon energies from 5 to 10 GeV.
Because we have integrated over $\omega$ we now use $\Gamma$ rather
than $\chi$ to label the curves.
We see in Fig. 2 the same structure as in Fig. 1, and we may draw similar
conclusions.  We also note that integration over these angles represents
$\sim 20\%$ of the gluon events.

\subsubsection{``Angular ordering'' effects}
As is well-known,
gluon emission off colourless $\qq$ pairs exhibits
so-called angular ordering behavior (see for example Refs. [\book,\webber]
and the discussion above
in Section 2):
If we split the radiation into pieces associated with the quark or the
antiquark, then integrate over the azimuthal angle about, say, the
quark's direction of motion, the quark piece of the radiation vanishes
for polar angles greater than the $q$--$\bar q$ angle.
(In particular this implies that {\it all} radiation is suppressed for
collinear $q$ and $\bar q$.)
This suppression is due to interference between gluons radiated by the
$q$ and the $\bar q$. A derivation of this result is given in
Appendix~E.

Because the width controls the
interference in the case of top, we
expect to see angular ordering behavior (or not)
according to the size of $\Gamma$.
To explore angular ordering for radiation off $b$'s from top decay,
we will examine the gluon emission probability integrated over
the azimuthal angle $\phi$ with respect to the $b$ direction for fixed
values of $\t12$.  The results for $\t12=5\degree,\ 30\degree,\ 90\degree
\
{\rm and} \ 150\degree$
are shown in Figures 3(a), 3(b), 3(c), and 3(d), respectively.

Before considering width effects, let us discuss the general features
of these curves; for this purpose Fig. 3(c) is the most instructive.
The distribution is shown as a function of $\theta$, the
gluon's polar angle with respect to the $b$ direction.
First we see, near $\theta=0$, the dead cone characteristic of emission off
heavy quarks:  emission is suppressed along the $b$ direction but
peaks nearby at the \lq dead-cone' angle,
 $\theta_c \sim m_b/E_b \approx 6\degree$ for $m_t = 140 \ \GeV$.
There is something like a dead cone at $\theta=\t12=90\degree$,
due to the $\bbar$.  The reason for the asymmetry between the $b$ and $\bbar$
is that we have integrated about the $b$ azimuthal
angle, but
we did not separate out the $\bbar$ radiation.
A final general feature we note is that most of the radiation is
in the vicinity of the quarks, so the distribution
falls off at very large $\t12$.

Now we examine how the width affects angular ordering.
In Fig. 3(a), $\t12=5\degree$.
The azimuthally integrated distribution is shown for $\chi=0$
(solid line), $0.5$ (dotted line), and $1$ (dashed line).
With the $b$ and $\bbar$ so close together, there are not two distinguishable
dead cones but one broadened peak.
As $\chi$ increases from 0 we see suppression of the radiation by the
interference, and for $\chi=1$ the emission is nearly eliminated; recall
that for collinear $q$ and $\bar q$ we expect no radiation at all.
As large as this effect is, however, it is of academic interest only:
because $b$ jets have a finite angular spread we can never hope to identify
events with this configuration.

In Fig. 3(b) we consider a larger $\bb$ angular separation,
$\t12=30\degree$; we can just begin to
discern the effect of the $\bbar$ dead cone.
We see again that as $\chi$ increases from 0 and the interference
turns on,
emission at angles larger than $\t12$ is suppressed.  We get maximum
suppression --- as much as an order of magnitude ---
for the coherent case, $\chi=1$.
Between the $b$ and $\bbar$, that is, for $\theta < \t12$,
the width makes no visible difference, but at larger angles the interference is
destructive.

In Figure 3(c) we show the same distribution for $\t12=90\degree$.
The effect appears less dramatic:
for any given $\theta_1$ the difference between $\chi=0$ and $\chi=1$
is not very large.  However, on closer inspection we notice
that the curves cross at $\theta=\t12$.  The radiation
outside the $b$ and $\bbar$ ({\it i.e.} $\theta>\t12$) is again suppressed,
but now the radiation between the $b$ and $\bbar$ is  enhanced,
so that the suppression of radiation outside the $\bb$ pair {\it relative}
to the radiation between is larger than it appears at first.
The net interference for this larger $\t12$ is constructive.

We consider a nearly back-to-back $\bb$ pair in Fig. 3(d) where
$\t12=150\degree$.  For such a large angular separation there is little
room outside the $\bb$ pair and the entire angular ordering effect amounts
to an enhancement of radiation between the $b$ and $\bbar$. This is the
well-known \lq string' [\lund] or \lq drag' [\drag] effect.

As an aside, we recall that all of the effects we discuss are also relevant
to photon radiation in the $\ww$ case.
A more complete numerical treatment of this will be given elsewhere [\wwpaper],
but in the meantime we give
an angular ordering example for
illustration.  We show in Figure 4 the $\ww$ analogues of
Figs. 3(b) and (d), for $\chi=0$ (solid lines) and 1 (dashed lines).
It should be noted that we have not included the effects of initial state
radiation, so that this figure corresponds, {\it e.g.,} to
$\gamma\gamma \to \ww$ rather than to $\ee\to\ww$.
Now because the lepton is nearly massless, the
radiation peaks are much sharper than for the $b$'s, but otherwise we
see similar features:  suppression outside and enhancement between the
$l^-$ and $l^+$ for $\theta_{12}$ large and small, respectively.  However,
there is one other important difference:  because of the large $W$ width
(2 GeV) and lower accessible photon energies, the relevant values of $\chi$
reach the width-sensitive range.  Furthermore, leptons do not hadronize,
so small lepton-antilepton angles are accessible.
Given sufficient event numbers, these
effects should be observable.

Returning to top, we have seen in Fig. 3, then,  that
the overall effect of the width on these azimuthally integrated distributions
is enhancement between the $b$ and $\bbar$ and suppression outside them.
This suggests another way to get at the width dependence experimentally:
look at the net radiation between and/or outside the $\bb$ pair.  We show
in Figures 5(a) and 5(b) distributions integrated over $\phi$ as in
Fig. 3, over gluon energies (again from 5 to 10 GeV)
and also over $\theta$.  In Fig. 5(a) we integrate $\theta$
from 0 to $\t12$ -- for the radiation between the $b$ and $\bbar$ --
and in Fig. 5(b) from $\t12$ to $\pi$ -- for the radiation
outside.

In Fig. 5(a) we see the enhancement of radiation between the $b$ and $\bbar$
as the width increases, and, as suggested in Fig. 3, the effect increases
with increasing $\t12$.  The radiation outside the $\bb$ pair shown in
Fig. 5(b) exhibits suppression of emission as the width increases.
The largest effect here is at small $\t12$; however we cannot hope to
do a measurement at very small angular separations because, again, $b$ jets
have finite angular size.  In both of these figures we see, as we did in
Fig. 2, that
the Standard Model width for $\mt=140\ \GeV$, 0.7 GeV (dotted line) gives
distributions very close to those for independent emission, and
that sensitivity to $\Gamma$ is only obtained in the few GeV range.

\subsubsection{Integrated distribution}
Finally we integrate over all angles and gluon energies from 5 to 10 GeV
to show the total soft gluon emission probability as a function of
$\t12$ in Figure 6.  As we shall see below when we discuss event rates, this
may be the only distribution we have very much hope of seeing
without multi-decades of collider runs.
The independent emission case, $\Gamma=0$, is, as always,
completely independent of $\t12$ due to the absence of interference.
Increasing the width turns on the interference and induces $\t12$
dependence.  The interference is destructive at small $\t12$ and
constructive at large $\t12$, and we see the crossover point
$\Tcr\approx 75\degree$ discussed in Appendix~B.  Also
evident here is the complete suppression of all radiation
for the coherent case (large $\Gamma$) when the $b$ and $\bbar$ are collinear.
And, again, there is sensitivity to $\Gamma$ as it approaches
the few GeV range, {\it i.e.} as it becomes comparable to the gluon energies
we consider.
Finally, note that the gluon emission probabilities are of the order
of $20\%$.

\subsection{Event rates}
This brings us to a discussion of event rates and the prospects for measuring
the top width from soft gluon distributions.  We must re-emphasize
that our results are at the parton level, and any realistic assessment must
incorporate hadronization of the $b$'s and gluons as well as
detector resolutions and acceptances, etc.
Having said that, we now look at cross sections.  The
 cross section
for $\tt$ production near threshold for $\mt=140\ \GeV$ is about 1 pb [\study].
If we assume a yearly luminosity of 10 fb$^{-1}$, this implies
10$^4$ $\tt$ events per year.  We saw in the previous subsection that the
number of events with a soft gluon is roughly 20\% of the lowest order rate,
or 2000 events.  If we further require leptonically ($e$ or $\mu$)
decaying $W$'s, we
are reduced to about 100 events/year.  This is less than promising, but
not quite hopeless.  If there is no other viable way to measure the top
width, soft gluon distributions may be an option.

\section{Conclusions}

How well does measuring the top width from soft gluon radiation
in $\ee\to\tt$ compare
with the standard technique of
scanning the threshold structure
of the total cross section?
Each method has its disadvantages.  The threshold structure is subject to
 uncertainties from beamstrahlung and
beam energy spread, and from theoretical higher order corrections and
dependence on parameters like $m_t$ and $\alpha_s$. The soft gluon radiation
method avoids these problems,
 but it is a higher order process with a lower event rate.
The two methods could therefore be considered as complementary --
the threshold cross section loses sensitivity with increasing width, but as we
have seen,
the gluon radiation pattern becomes {\it more} sensitive at
larger $\Gamma$ for accessible gluon energies.
For most of the expected top
mass range, the threshold structure  method {\it is} probably better, but if
$m_t$ and $\Gamma$ are large, then examining  soft gluons may be more useful.

In summary, we have seen that the top quark's large width gives rise to
new effects from the interplay between the strong and weak interactions,
and that the top width affects the distributions of soft gluons radiated
in top events.  Near
the $\tt$ threshold, the effect of the width is to suppress the
{\it interference} between gluons radiated by the $b$ and $\bbar$,
in contrast to the expectation of coherent radiation from the $\bb$ pair.
The sensitivity of the gluon distribution to $\Gamma$ is largest
for gluons with energy $\omega \sim\Gamma$. If 5 GeV corresponds to a realistic
minimum energy for measurable gluon jets, then the Standard Model width
 of a 140 GeV top quark is sufficiently smaller than this to almost
completely suppress the interference.

Note that the results of our analysis could in principle be incorporated
into a Monte Carlo scheme for generating final states in
$\ee\to\tt\to W\pw+b W\pw- \bar b$ events. Contrary to the standard
expectation (see for example [\fujii]), the $\bb$ antenna is practically
inactive here since the bulk of the radiation, that is primary gluons
with $\omega > \Gamma$ (in the top rest frame), is governed
by the $t\bar b$ and $\bar t b$ antennae, and is thus unaffected
by the relative $\bb$ orientation angle  $\theta_{\bb}$.
When parton cascades are included
in the picture, the corresponding hard scale $Q$ is given by
\beq
 Q \sim E_b \approx {M\pw2 -M_W\pw2 \over 2  M}   \; .
\eeq
The only particles which {\it are} sensitive to $\theta_{\bb}$
are those originating from primary brems\-strah\-lung
gluons with $\omega \lapproxeq \Gamma$, whose yield is determined
by the parton cascade scale $Q\sim \Gamma$.

Finally, we have also extended our analysis to the case of photon radiation
near threshold in $\ee\to\ww$, including both hadronic and leptonic
$W$ decays. This is especially relevant for LEP200, where the measurement
of soft photons with $\omega \sim \Gamma_W$ would also reveal interesting
interference effects. We will present numerical
results for  this elsewhere [\wwpaper].

\bigskip
\medskip
\noindent{\Large\bf Acknowledgements}

\bigskip
\noindent
This work was supported in part by the Texas National Research Laboratory
Commission and the United Kingdom Science and Engineering Research  Council.
Yu.L.D. and L.H.O thank the Center for Particle Theory at the University of
Durham for hospitality while some of this work was being completed.
We are grateful to D. Borden, V. Gribov, P. M\"attig, M. Perl and M. Swartz for
useful discussions.

\newpage
\appendix
\setcounter{equation}{0}
\renewcommand{\theequation}{\Alph{section}\arabic{equation}}

\section{Semi-classical derivation of the radiation
pattern}

The decay of a heavy $t$ quark
at rest produces a fast-moving $b$ quark and thus causes acceleration
of the colour charge. We are interested in gluon bremsstrahlung
induced by this acceleration. Analogously to the treatment of
classical electromagnetic currents [\oldqed], the colour field
formation is conveniently described in terms  of
Lienard-Wiechert potentials. Thus the
two quark currents which participate in the colour field formation are:
\bmini
\vec{j}_1&=&\Vo \>\delta\pw3(\vec{r}-\Vo\>(t-t_{01}))
\cdot\vartheta(t-t_{01})\>,\\
\vec{j}_2&=&\Vt \>\delta\pw3(\vec{r}-\Vt\>(t-t_{02}))
\cdot\vartheta(t-t_{02})\>,
\emini
where $t_{i0}$ are the times of the two decays.
The emission amplitude for the field component with the 4-momentum
$(\omega,\vec{k})$ is proportional to the Fourier transformed total current,
which we write introducing the effective ``colour charge'' as
\beq\label{jtot}
 \vec{j}(k)=\sqrt{g_s\pw2\,C_F}\>\left(\vec{j}_1(k) - \vec{j}_2(k)\right)\>,
\eeq
where the relative minus sign reflects the opposite charges,
and $g_s^2 = 4\pi\alpha_s$.
For each of the two terms of (\ref{jtot}) we have
\beeq
\vec{j}_i(k) &=& \int_{-\infty}\pw\infty dt\int d\pw3r\>
e\pw{i\, x\pw{\mu} k_{\mu}}\>\vec{j}_i(t,\vec{r})
\nonumber \\
\label{jik}
  &=& \vec{v}_i\frac{i}{k\pw0- (\vec{k}\cdot\vec{v}_i)}
\cdot e\pw{i k\pw0 t_{0i}}\>.
\eeeq
The field potential induced by the current (\ref{jik}) at large
(positive) time $x\pw0$ reads
\beq\label{pot1}
\vec{A}_i(x) \>=\> \int\frac{d\pw4 k}{(2\pi)\pw4}\> e\pw{-ix\pw{\mu}
k_\mu}\>[-2\pi i\delta(k\pw2)]\cdot \vec{j}_i(k)
\ =\  \int\frac{d\pw3 k}{2\omega (2\pi)\pw3}\>
e\pw{-i\omega x\pw{0}+i(\vec{k}\cdot\vec{x})}\cdot \vec{A}_i( {k})\> .
\eeq
The momentum Fourier component of the total vector field  is given by
\bminiG{potres}
\label{pott}
 \vec{A}( {k}) &=&  \sqrt{g_s\pw2\,C_F}\>
\left(  \vec{A}_1( {k}) -  \vec{A}_2( {k})  \right)\>; \\
\label{potk}
 \vec{A}_i( {k}) &=&  \frac{\vec{v}_i}{\omega(1-v_i\cos\theta_i)}\cdot
e\pw{i\omega t_{0i}}\>;\qquad \omega= |\vec{k}| \>,
\>\> (\vec{k}\cdot\vec{v}_i)=\omega v_i\,\cos\theta_i\>.
\emini
To calculate the radiation probability we square the projections of the full
field amplitude (\ref{pott}) onto two ``physical'' gluon states
$\vec{e}_{\lambda}$,
where $(\vec{e}_{\lambda}\cdot\vec{k})\!=\!0$,
$(\vec{e}_{\lambda})\pw{2}\!=\!1$,
and sum over polarizations to obtain
\beq\label{radprob}
dN = \frac{d\pw3k}{2\omega(2\pi)\pw3}
\sum_{\lambda=1,2}\abs{\vec{A}(k)\cdot
 \vec{e}_{\lambda} }\pw2
=\frac{d\pw3k}{2\omega(2\pi)\pw3}
\sum_{\alpha,\beta=1,2,3} \>A\pw{\alpha} (k)\cdot
\left[\delta_{\alpha\beta}-\frac{k_\alpha\,k_\beta}{\vec{k}\pw2}\right]
\cdot A\pw{*\beta} (k) \>.
\eeq
Substituting for the vector field from (\ref{potres}) and making
 use of the relations
\bminiG{esum}
v_{i\alpha}
\left[\delta_{\alpha\beta}-\frac{k_\alpha\,k_\beta}{\vec{k}\pw2}\right]
v_{i\beta} &=& v_i\pw2\sin\pw2\theta_i\>,\\
\label{esumint}
v_{1\alpha}
\left[\delta_{\alpha\beta}-\frac{k_\alpha\,k_\beta}{\vec{k}\pw2}\right]
v_{2\beta} &=& v_1v_2(\cos\theta_{12}-\cos\theta_1\cos\theta_2)\>,
\emini
we conclude that the $\abs{A_1}\pw2$ and $\abs{A_2}\pw2$ terms reproduce the
sum of the ``independent'' radiation contributions $\cRi$
(\ref{indep}) while the
interference $2\Re(A_1\,A_2\pw*)$ is proportional to $2 \cJ$ (\ref{coher}).
So finally we arrive at
\beq\label{deltat}
dN=\frac{d\omega}{\omega}\frac{d\Omega}{4\pi}\>\frac{C_F\as}{\pi}\cdot
\left\{ \cR_1 + \cR_2 +2\Re e\pw{i\omega(t_{01}-t_{02})}\cdot \cJ\right\}\>.
\eeq
This expression describes the radiation accompanying the process with
heavy top quarks decaying at times $t_{0i}$ after the $\tt$ production.
These times are not measured, but are distributed according to the
decay exponentials
\beq
\left[\>\Gamma\int_0\pw\infty\>dt_{0}\,e\pw{-\Gamma t_{0}}\>\right]\>.
\eeq
Substituting (\ref{deltat}) into the decay-time integrals we see that the
interference term gives the $\chi$ factor,
\bminiG{gamav}
\lrang{e\pw{\pm i\omega t_{0i}}}_i &\equiv&
\Gamma\int_0\pw\infty\>dt_{0i}\,e\pw{-\Gamma t_{0i}}
\cdot e\pw{\pm i\omega t_{0i}}
=\frac{\Gamma}{\Gamma\mp i\omega}\>,\\
\label{getchi}
\lrang{\lrang{e\pw{\pm i\omega( t_{01}-t_{02})}}_1}_2 &=&
\frac{\Gamma}{\Gamma+ i\omega}\>\frac{\Gamma}{\Gamma- i\omega}=\chi(\omega)\>,
\emini
 leading to the final
expression which is identical to the representation (\ref{splitJa}).

Thus we conclude that the $\omega$-dependence of the soft radiation is due
to incoherence induced by the uncertainty $\Delta t_0\sim\Gamma\pw{-1}$ in the
acceleration times of the two ($b$-quark) charges.
Such a delay can be resolved by a gluon with a {\em small}\/ wavelength
$$
\lambda\sim\omega\pw{-1}\>\la\> \lrang{\Delta t_0}\sim \Gamma\pw{-1}
$$
in which case ($\omega\ga\Gamma$) the coherence gets lost and the radiation
pattern reduces to the sum of the two independent $b$ and $\bar{b}$
contributions, {\it i.e.}\/ $\widehat{tb}$
and $\widehat{\bar{b}\bar{t}}$ antennas, (see (\ref{ind})).
On the other hand, for  wavelengths {\em large}\/
compared to $\Gamma\pw{-1}$, the (\ref{coh}) regime,
the time delay does not affect the radiation:
coherence remains undisturbed and the pattern is given by the
$b\bar{b}$ antenna describing the point-like production of the ``light''
quark pair,
just as if  there was no $t\bar{t}$ stage at all.
The two situations can be represented pictorially as shown below.

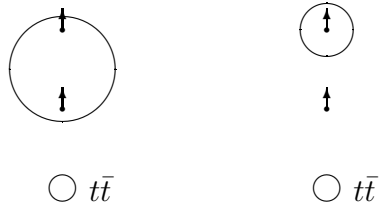
\begin{center}
\begin{picture}(200,80)
\newsavebox{\two}
\savebox{\two}{
\put(0,0){\circle{10}}
\put(10,-5) {$t\bar{t}$}
\put(0,30){\vector(0,1){8}}
\put(0,30){\circle*{2}}
\put(0,60){\vector(0,1){8}}
\put(0,60){\circle*{2}}
}
\put(0,0){\usebox{\two}}
\put(0,45){\oval(40,40)}
\put(100,0){\usebox{\two}}
\put(100,60){\oval(20,20)}
\end{picture}
\end{center}

\noindent
Note that the result (\ref{getchi}) for the decay
profile function can easily be generalized
to the case of {\em different}\/ decay widths.
This arises, for example, in the case of the production and decay
of a pair of different supersymmetric particles, for example
$\tilde{q}\tilde{g}$, where the decay width of the decaying
particles could in principle be very different.
In the general case we obtain
\beq\label{chigen}
\chi(\omega)= \Re
\left\{\frac{\Gamma_1}{\Gamma_1+ i\omega}\>\frac{\Gamma_2}{\Gamma_2- i\omega}
\right\}
= \frac{\Gamma_1\Gamma_2(\Gamma_1\Gamma_2+\omega\pw2)}
{(\Gamma_1\pw2+\omega\pw2)(\Gamma_2\pw2+\omega\pw2)} \>.
\eeq
For example, for very  different decay times, say $\Gamma_1\gg\Gamma_2$
(and $\omega\sim\Gamma_2$)
the expression (\ref{chigen}) would lead us back to the original
$$
 \chi(\omega) = \frac{\Gamma_2\pw2}{\Gamma_2\pw2+\omega\pw2}\>.
$$
However we note, in this context, that our perturbative treatment
of gluon emission only makes sense if $\omega > \mu \sim 1\; \fm\pw{-1}$.
Unfortunately this means we cannot discuss in this way the interesting
cases of charged Higgs decay $H\pw+ \to t \bar{b}$ or single
top production $g W\pw+ \to t \bar{b}$ where $\Gamma_b \ll \mu$
and $\chi \ll 1$.

\newpage
\setcounter{equation}{0}
\section{Angular-integrated distributions in $\ee\to\tt$}

Using the result
\begin{equation}
\int_{-1}\pw{1}d\cos\theta\>\frac{v\pw2\,\sin\pw2\theta}{(1-v\cos\theta)\pw2}
\>=\> \frac{2}{v} \ln\frac{1+v}{1-v} -4\>,
\end{equation}
we see that the angular-integrated contributions of the independent terms
are
\begin{equation}
\int\>\frac{d\Omega}{4\pi}\>\cRi
\>=\> \frac{1}{v_i} \ln\frac{1+v_i}{1-v_i} -2\> \> \equiv \cI(v_i)\>,
 \quad (i=1,2)\> .
\label{intcri}
\end{equation}

The coherent contribution (\ref{coher}) contains the
integral
\beq
\label{beint}
\int \frac{d\Omega}{4\pi}
\>\frac{2(1-v_1v_2\cos\theta_{12})}{(1-v_1\cos\theta_1)(1-v_2\cos\theta_2)}
=\int \frac{d\Omega}{2\pi}\>\frac{\omega\pw2\>(p_1p_2)}{(p_1k)\,(p_2k)}
= \frac{1}{r} \>\ln  \frac{1+r}{1-r} \>;
\eeq
where
\beq
\label{rdef}
 r \>\equiv\>
\sqrt{1-\frac{(1-v_1\pw2)(1-v_2\pw2)}{(1-v_1v_2\cos\theta_{12})\pw2}}
= \sqrt{1-\frac{M_1\pw2\,M_2\pw2 }{(p_1p_2)\pw2}} \>,
\eeq
the Lorentz-invariant quantity that is closely related to the relative
quark velocity in
the rest frame of the  pair.
In terms of the invariant energy $s=(p_1 + p_2)\pw2$ and the c.m.s. momentum
\beq
p_c\pw2 = \frac{\left[s-(M_1\!+\!M_2)\pw2\right]
\left[s-(M_1\!-\!M_2)\pw2\right]}{4\>s}
\eeq
one has
\beq
 r \>=\> \frac{2p_c\>\sqrt{s}}{s-(M_1\pw2+M_2\pw2)}\>.
\eeq
For the case of equal masses, $M_1\!=\!M_2$, (\ref{rdef})
is related to the quark c.m.s. velocity $v_c$ by
\beq
 r = \frac{2v_c}{1+v_c\pw2}\>.
\eeq
Combining (\ref{beint}) with the two remaining terms of (\ref{coher})
which give a  constant subtraction, we obtain (\cf (\ref{intcri}))
\beq\label{cRcty1}
\int\>\frac{d\Omega}{4\pi}\>\cRc=
\frac1r \>\ln\frac{1+r}{1-r} -2 \>=\> \cI(r)\>.
\eeq

Putting everything together gives, for the angular integrated
radiation yield,
\beeq
\frac{dN}{d\omega} \> &=&\> \frac{C_F\alpha_s}{\pi\omega}\>
\left\{ (1-\chi(\omega)) \left[ \frac{1}{v_1}\log\frac{1+v_1}{1-v_1}
+  \frac{1}{v_2}\log\frac{1+v_2}{1-v_2} -4 \right] \right. \nonumber
\\
&&\left.  \> + \chi(\omega) \left[ \frac{1}{r}\log\frac{1+r}{1-r} -2
 \right] \right\} \> .
\label{inttot}
\eeeq
We see that
the total radiation splits into incoherent and coherent contributions,
\beq
\frac{dN}{d\omega} = \frac{C_F\as}{\pi\omega }
\left\{\> (1\!-\!\chi(\omega)) \cdot \left[\> \cI(v_1) + \cI(v_2)\>\right]
\>+\> \chi(\omega) \cdot \cI(r)\>\right\}\>,
\eeq
the relative weight of which is controlled by the profile function $\chi$
which depends on the $\omega/\Gamma$ ratio.
The first contribution consists of two $\cI(v_i)$ terms describing
independent radiation
off quark antennae ``attached'' to the c.m.s. ({\it i.e.} in the rest frame
of the decaying top quarks).
The  function $\cI$ can be expressed in terms of the
``4--angle'' $\eta$ of the quark momentum as
\beq
\cI(v_i)= 2\left(\>\frac{\eta_i}{\tanh(\eta_i)} -1\>\right)\>; \quad
v_i=\tanh(\eta_i)\>.
\eeq
The coherent contribution, in contrast, carries
no information about the initial
$\tt$ system but depends exclusively on the relative motion of the two
final colour charges, the
$\widehat{(12)}$  antenna.
The argument of the corresponding $\cI$ factor here is the relative
``4--angle'' between the quarks,
\beq
\cI(r)= 2\left(\>\frac{\Delta}{\tanh(\Delta)} -1\>\right)\>;
\quad  r=\tanh(\Delta)\>,
\eeq
where $\Delta=\eta_1+\eta_2$ has to be calculated in a reference frame
where \qq are
{\em anti-collinear} (\eg in the c.m.s. of the pair).

In practice, we are usually working in the ultra-relativistic limit
$(1-v_i ) \ll 1$, where Eq. (\ref{inttot}) is dominated by the logarithmic
collinear singularities.
Since the angular integral of the interference term $\cJ$ (Eq. (\ref{coher}))
converges at $v_1=v_2=1$,
one would expect the main collinear contributions $\propto \log(1\!-\!v_i)$
to be $\omega$-independent.
To verify this let us keep only the
non-vanishing logarithmic and constant terms
in (\ref{inttot}), neglecting powers of $(1 - v_i) \ll 1$.
We can approximate (\ref{rdef}) as
\beq\label{r2app}
r\pw2 \approx 1- \frac{ 4\,(1-v_1) (1-v_2)}
{\left[\>1-\cos\theta_{12} + [(1\!-\!v_1)+(1\!-\!v_2)]\cos\theta_{12}
\>\right]\pw2}\>.
\eeq
Thus for  $\theta_{12}$ values not particularly  close to 0,
so that we can neglect the second term in the denominator
compared to the first,
we obtain
\beq\label{rapp}
1- r \>\approx\> 2\cdot \frac{ (1-v_1)
(1-v_2)}{(1-\cos\theta_{12})\pw2} \ll 1\>,
\eeq
and the expression in curly brackets in (\ref{inttot}) becomes
\beeq
\label{tyappr}
\{\>\}
\ &\approx&\ \left[\ln\frac{2}{1\!-\!v_1} + \ln\frac{2}{1\!-\!v_2} -4 \right]
+ 2\chi(\omega) \left[ \>\ln\frac{1-\cos\theta_{12}}{2} + 1 \right]\>.
\eeeq
Thus we have verified that the main ``collinear'' contributions are
$\omega$-independent
in the  region of relative quark angles  $\theta_{12}$
{\em exceeding}\/ the aperture of the corresponding ``dead cones''.
It is interesting to compare the relative size
of the two terms in (\ref{tyappr}) in practice. For example, for the
case of a 140~GeV top quark pair we find
$$
\{\>\}
\ \simeq\ 7.9\
+ \ 2\chi(\omega) \left[ \>\ln\frac{1-\cos\theta_{12}}{2} + 1 \right]\>,
$$
showing that the integrated quantity is indeed quite sensitive
to $\chi$ and hence to the width.
(In contrast, for a $WW$ pair decaying to muons and neutrinos the
numerical term from the logarithms is 23, and the sensitivity
to $\Gamma$ is much decreased.)
One could imagine,
for example, measuring the profile function $\chi$ by studying the
$\theta_{12}$ variation of the total radiation yield.

Note in particular that the second $\omega$-dependent
term in (\ref{tyappr})
{\em enhances}\/ the radiation at large $\theta_{12}$ and acts
{\em destructively}\/ when $\theta_{12}$ is chosen below the value
$\Theta_{\mbox{\tiny crit}}$ given by
\beeq
\label{Tcrcos}
\cos\Tcr &=&  1-2\exp(-1)  = 0.26424 \>,  \nonumber \\
\Tcr &    \approx &  75\pw0\>.
\eeeq

It is interesting to notice that the suppression at
$\theta_{12}\to 0$ can be strong enough to completely compensate
the        main
collinear contributions.
Indeed, taking parametrically small angles
$$
 \theta_{12}\pw2 \>\ll\> (1-v_1) \>+\> (1-v_2)\>,
$$
we would obtain for $r$ the value (see (\ref{r2app}))
\beq
r\pw2 \> \approx \>
 \frac{(v_1-v_2)\pw2}{\left[\> (1\!-\!v_1)+(1\!-\!v_2)\>\right]\pw2}
\eeq
which can be arbitrarily small for nearly equal quark velocities.
If $r\ll 1$ the second (coherent)  term in the general expression
(\ref{inttot})
for the radiation yield becomes negligible
and one is left with ($v\equiv v_1\!\approx\!v_2$)
$$
\frac{dN}{d\omega} \>\approx\> 2\>  \frac{C_F\as}{\pi\omega }\cdot
(1\!-\!\chi(\omega)) \ \left[\>\frac{1}{v}\ln\frac{2}{1\!-\!v} -2\> \right] \>.
$$
The result vanishes when $\chi(\omega)\to1$ (\ie for $\omega\ll\Gamma$).
This corresponds to the total coherent suppression of radiation off two
opposite
charges moving in the same direction with equal velocities.
Notice that for $\omega=\Gamma$ the independent radiation off collinear
daughter particles is suppressed by a factor 2.

It is worth mentioning that
the $\omega$-dependent coherent effect cancels after integration over
all angles $\theta_{12}$.
This means that the interference $\cJ$
does not affect the {\bf total} bremsstrahlung
caused by the decay of the heavy unstable
objects,
but only redistributes the accompanying radiation between
configurations with different relative angles $\theta_{12}$.
This
can be checked explicitly by evaluating the angular integral of the
second (coherent) term of
(\ref{inttot}) over $\theta_{12}$ which results in
\beq
\eqalign{
& \int_{0}\pw{\pi}  \sin\theta_{12} d\theta_{12}\int\frac{d\Omega}{4\pi}\>\cRc
= \int_{-1}\pw{+1} d\cos\theta_{12} \>
\left[\>\frac1r\ln\frac{1+r}{1-r}-2\>\right] \cr
&= 2\cdot\left\{\frac1{v_1}\ln\frac{1+v_1}{1-v_1} +
\frac1{v_2}\ln\frac{1+v_2}{1-v_2}  -4 \right\}
\equiv \int_{0}\pw{\pi} \sin\theta_{12}d\theta_{12}
\cdot \int\frac{d\Omega}{4\pi}\>\cRi\>.
}\eeq

This suggests a way of extracting the
profile function $\chi(\omega)$ by
studying the coherent ``redistribution'' effects in the total radiation yield.
This could be  done for example by comparing the gluon yield at fixed
$\theta_{12}$ below and above the $\Tcr$ value.

Let us consider, therefore,
the integrated quantity characterising the interference effects,
namely the difference of the integrals over the $b\bar{b}$ opening angles
{\em above}\/ and {\em below}\/  $\Tcr$:
\beq\label{Ddef}
 \delta \equiv
\frac{1}{(1+\cos\Tcr)}\int_{\Tcr}\pw{\pi}\sin\theta_{12}d\theta_{12}\>
\frac{\omega dN}{d\omega}
-\frac{1}{(1-\cos\Tcr)}\int_{0}\pw{\Tcr}\sin\theta_{12}d\theta_{12}\>
\frac{\omega dN}{d\omega}\>.
\eeq
The normalization in (\ref{Ddef}) is chosen so that the
$\theta_{12}$--{\em independent}\/ contributions cancel.
Within the relativistic approximation, we have for (\ref{tyappr})
\beq\label{cNtt}
\cN=\cRi +2\chi(\omega)\cdot\left[\>\ln\frac{1-\cos\theta_{12}}{2} + 1\>\right]
\>.
\eeq
Making use of the definition of the critical angle (\ref{Tcrcos}),
we calculate the integrals
\bmini
\frac{1}{(1+\cos\Tcr)}\int_{-1}\pw{\cos\Tcr}d\cos\theta_{12}\>\left[\>
\ln\frac{1-\cos\theta_{12}}{2} + 1\>\right] &=&
\frac{1-\cos\Tcr}{1+\cos\Tcr} \>,\\
\frac{1}{(1-\cos\Tcr)}\int_{\cos\Tcr}\pw{1}d\cos\theta_{12}\>\left[\>
\ln\frac{1-\cos\theta_{12}}{2} + 1\>\right] &=& -1\>;
\emini
to obtain finally for (\ref{Ddef})
\beq\label{chimes}
\delta = \frac{C_F\as}{\pi}\>\frac{4}{1+\cos\Tcr}\cdot \chi(\omega)\>.
\eeq
Invoking the numerical value of the critical angle (\ref{Tcrcos}) we write
 (\ref{chimes}) as
\beq\label{Dfin}
 \delta
\>=\> 4.218 \>\frac{\as(\omega)}{\pi}\cdot\chi(\omega)\>.
\eeq
Notice that we have chosen  here the gluon energy as the argument
of the running coupling,
since it is {\it large} gluon energies that contribute to (\ref{Dfin}).


\newpage
\setcounter{equation}{0}
\section{Initial and final state radiation in
$e\pw+e\pw-\to W\pw+W\pw-$ near threshold}

Let us denote by $\Vz$ ($-\Vz$)  the 3-velocities of the incoming
$e\pw+$ ($e\pw-$).
Then the initial state  electromagnetic current takes the form
\beq
\vec{j}_0 =
\left[\> \Vz \delta\pw3(\vec{r}-\Vz\, t) - (-\Vz )
\delta\pw3(\vec{r}+\Vz\, t) \>\right] \cdot\vartheta(-t)
\eeq
which gives an extra contribution to the field amplitude  (see Appendix~A)
\beq\label{inampl}
\vec{A}_0 = -
 \frac{\Vz}{\omega} \left[\>\frac1{1-(\Vz\Vn)} +
\frac1{1+ (\Vz\Vn)} \>\right]= -
\frac{\Vz}{\omega}\,
\frac{2  }{1-v_0\pw2\cos\pw2\theta_0} \>,
\eeq
where $\Vn$ is direction of the photon and
$\theta_0$ its angle with respect to the incoming {\em positron}.
Since the $\ee$  ``disappear'' at the same
time as the $WW$ pair is produced,
the amplitude (\ref{inampl}) is real with our convention ($t\!=\!0$).
Let us choose for the sake of simplicity the $(\ell\nu\ell\nu)$ decay channel
of the $WW$.
We use (\ref{radprob}) to obtain for the radiation probability
\beeq
\label{asmall}
dN &=& e\pw2 \frac{d\pw3k}{2\omega(2\pi)\pw3}
\;  \sum_{\lambda=1,2}\abs{\vec{A}(k) \!\cdot\!
 \vec{e}_{\lambda} }\pw2
=\frac{e\pw2}{4\pi\pw2}
\,\frac{d\omega}{\omega}\, \frac{d\Omega_{\Vn}}{4\pi}
\!\!\sum_{\alpha,\beta=1,3} \!\! a\pw{\alpha} (k)\,
\left[\> \delta_{\alpha\beta}-n_\alpha\,n_\beta \>\right]\,
a\pw{*\beta} (k) \>,\nonumber \\
&& \vec{a}(k) =
 \frac{\Vo}{1-(\Vo\Vn)}\cdot e\pw{i\omega t_{01}}
\>-\>
 \frac{\Vt}{1-(\Vt\Vn)}\cdot e\pw{i\omega t_{02}}
-\frac{2\Vz}{1-(\Vz\Vn)\pw2}
\>,
\eeeq
where $\Vo$, $\Vt$ stand for positively and negatively charged final
state leptons respectively.
The novel feature of (\ref{asmall}) is an interference between the initial
state radiation (ISR) (the last term) and the final state radiation (FSR)
(the first two terms), which has the structure
\beq
 \cJ_{01}\cdot 2\Re e\pw{i\omega t_{01}} +
\cJ_{02}\cdot 2\Re e\pw{i\omega t_{02}}\>.
\eeq
After the integration over the $W\pw\pm$ decay times $t_{0i}$ is performed
(see (\ref{gamav})) it gives rise to the same profile function $\chi$, due to
the identity
\beq
\Re \left\{\>\frac{\Gamma}{\Gamma + i\omega}\>\frac{\Gamma}{\Gamma - i\omega}
\>\right\}
= \Re\left\{\> \frac{\Gamma}{\Gamma \pm i\omega}\>\right\}
= \frac{\Gamma\pw2}{\Gamma\pw2 +\omega\pw2} \equiv \chi(\omega) \>.
\eeq

The final answer can therefore be represented as
\beq
dN = \frac{\al}{\pi}\,\frac{d\omega}{\omega}\, \frac{d\Omega_{\Vn}}{4\pi}\>
\left[\>\cN_{\mbox{\tiny FS}} +\cN_{\mbox{\tiny IS}} +\cN_{\mbox{\tiny
I/F}} \>\right]
\eeq
where the first term
is, as before (\cf (\ref{splitJa})),
\bminiG{threeIF}
\cN_{\mbox{\tiny FS}} &=& \cR_1 + \cR_2 +  2\chi(\omega)\cdot
\frac{(\Vo\Vn)(\Vt\Vn)-(\Vo\Vt)}{[\,1-(\Vo\Vn)\,][\,1-(\Vt\Vn)\,]} \>.
\eeeq
The new terms describe independent radiation off the  initial state
$\ee$ antenna,
\beeq
\cN_{\mbox{\tiny
IS}}&=&\frac{4v_0\pw2\sin\pw2\theta_0}{(1-v_0\pw2\cos\pw2\theta_0)\pw2}\>,
\eeeq
and the ISR/FSR interference contribution
\beeq
\label{IFint}
\cN_{\mbox{\tiny I/F}} &=& 2\chi(\omega)\cdot \frac{2}{1-(\Vz\Vn)\pw2}
\left\{\> \frac{(\Vz\Vn)-(\Vz\Vo)}{1-(\Vo\Vn)} -
\frac{(\Vz\Vn)-(\Vz\Vt)}{1-(\Vt\Vn)} \>\right\}\>.
\emini
Note that the ISR/FSR
interference (\ref{IFint})  vanishes (i)  after integration
over the angles between the ISR and FSR antennae
(keeping the relative angle between the daughter charged particles fixed),
and (ii) in the limit $\omega \gg \Gamma$, as expected.
We recall also that both $\cN_{\mbox{\tiny IS}}$
and $\cN_{\mbox{\tiny I/F}}$
vanish when the kinematic limit
$\omega > \omega_{max}\pw{\mbox{\tiny IS}}$
(Eq. (\ref{omegamax})) is exceeded.

In practice we are not interested in photon emission close to
the beam direction, and so for $\theta_0 \gg m_e/M_W$ we can set
$v_0 = 1$   and the above expressions become
\beeq
\cN_{\mbox{\tiny IS}}&\simeq&\frac{4}{\sin^2\theta_0}\>,
\eeeq
and
\beeq
\label{IFinta}
\cN_{\mbox{\tiny I/F}} &=& \chi(\omega)\cdot
\cN_{\mbox{\tiny IS}} \cdot
\left\{\> \frac{\cos\theta_0-(\Vz\Vo)}{1-(\Vo\Vn)} -
\frac{\cos\theta_0-(\Vz\Vt)}{1-(\Vt\Vn)} \>\right\}\>.
\eeeq

\newpage
\setcounter{equation}{0}
\section{Hadronic $W^+W^-$ decays}

In this Appendix we derive the expressions for the photon radiation
pattern in $e^+e^-\to W^+W^-$ when at least one of the $W$'s
decays {\it hadronically }, {\it i.e.} to $q\bar{q}'$.
Consider first the case when both $W$'s decay hadronically. The two
currents are then (see Eq.(\ref{twocurrH}))
\begin{eqnarray}
j_1^\mu &=&  Q\,\,\frac{p_1^\mu}{(kp_1)} +
(1-Q\,)\,\frac{\bar{p}_1^\mu}{(k\bar{p}_1)}
- \frac{q^\mu}{(kq)} \>, \nonumber \\
j_2^\mu &=& Q'\,\frac{p_2^\mu}{(kp_2)} +
(1-Q')\,\frac{\bar{p}_2^\mu}{(k\bar{p}_2)}
- \frac{q^\mu}{(kq)} \>,
\end{eqnarray}
where $Q=Q'=\frac{2}{3}$ and $p_1$, $p_2$ denote the momenta of the {\em
up}-type quark and antiquark
respectively.
In the 3-vector form ($(eq)=0$) we can write
\beq
 \vec{j} = \frac{\vec{v}}{\omega} \left[ \>
\frac{Q}{1-\Vn\vec{v}} - \frac{1\!-\!Q}{1+\Vn\vec{v}} \>\right]
= \frac{\vec{v}}{\omega} \> \frac{(2Q-1) + \Vn\vec{v} }{1-(\Vn\vec{v})^2}
\eeq
where we have treated the quark and the accompanying antiquark momenta as
{\em anti-parallel}\/ and of equal mass.
Then
\begin{eqnarray}
\cN^{(qq)} &\propto &  \left[ (e\Vo)\frac{(2Q-1 )+\Vn\Vo}{1-(\Vn\Vo)^2}
\right]^2
+ \left[ (e\Vt)\frac{(2Q'-1)+\Vn\Vt}{1-(\Vn\Vt)^2} \right]^2 \nonumber \\
& &- 2\chi(\omega)\cdot
(e\Vo)(e\Vt)\left[\frac{(2Q-1 )+\Vn\Vo}{1-(\Vn\Vo)^2}\right]\,
\left[\frac{(2Q'-1 )+\Vn\Vt}{1-(\Vn\Vt)^2}\right].
\label{Nqq}
\end{eqnarray}
For the mixed one-quark-one-leptonic decay
configuration, $Q'=1$   and we have
\begin{eqnarray}
\cN^{(q\ell)} &\propto &  \left[ (e\Vo)\frac{(2Q-1
)+\Vn\Vo}{1-(\Vn\Vo)^2} \right]^2 +
\left[ (e\Vt)\frac{1}{1- \Vn\Vt} \right]^2 \nonumber \\
& &- 2\chi(\omega)\cdot  (e\Vo)(e\Vt)\left[\frac{(2Q-1 )
+\Vn\Vo}{1-(\Vn\Vo)^2}\right]\, \left[\frac{1}{1- \Vn\Vt }\right].
\label{Nql}
\end{eqnarray}
Experimentally it seems very difficult (if at all possible) to discriminate
between the two jets that originate from, say, $W\pw+\to u+\bar{d}$ decay.
Without being able to {\em separate}\/ quark and antiquark jets ($Q$
from $1\!-\!Q$) we
have to drop in (\ref{Nqq},\ref{Nql})
the {\em odd}\/ terms in $2Q\!-\!1$ ($2Q'\!-\!1$).
After summing over photon polarizations, Eq.~(\ref{esum}), we arrive at
\bminiG{NqqlA}
\label{NqqA}
\eqalign{
\cN^{(qq)} \propto  &   v_1^2\sin^2\theta_1
\frac{(2Q\!-\!1 )^2+ v_1^2\cos^2\theta_1}{(1-v_1^2\cos^2\theta_1)^2 }
+  v_2^2\sin^2\theta_2
\frac{(2Q'\!-\!1 )^2+ v_2^2\cos^2\theta_2}{(1-v_2^2\cos^2\theta_2)^2 }    \cr
 &+ 2\chi(\omega)\cdot v_1v_2\,(\cos\theta_1\cos\theta_2-\cos\theta_{12})
 \left[\frac{ v_1 \,\cos\theta_1}{1-v_1^2\cos^2\theta_1} \right]\,
 \left[\frac{  v_2\,\cos\theta_2}{1-v_2^2\cos^2\theta_2} \right]  ;  }  \\
\label{NqlA}
\eqalign{
\cN^{(q\ell)} \propto &   v_1^2\sin^2\theta_1
\frac{(2Q\!-\!1 )^2+ v_1^2\cos^2\theta_1}{(1-v_1^2\cos^2\theta_1)^2 }
\>\>+\>\>  \frac{v_2^2\sin^2\theta_2}{(1-v_2\cos\theta_2)^2 }    \cr
 &+ 2\chi(\omega)\cdot v_1v_2 \,(\cos\theta_1\cos\theta_2-\cos\theta_{12})
  \left[\frac{ v_1\cos\theta_1 }{1-v_1^2\cos^2\theta_1} \right]\,
  \left[\frac1{ 1-v_2 \cos \theta_2}  \right] .  }
\emini
Recalling the expressions for $\cRone$, $\cRtwo$ and $\cJ$ introduced
in Section 2, it is straightforward to cast this result in the form given
in Eqs.~(\ref{Nqqeq},\ref{Nqleq}).
Note that there is a subtlety in the $\cN^{(qq)}$ case concerning the
definition of the angles. Whereas for the $\tt$ case we could
unambiguously define, say, $\theta_1$ with respect to the $b$ quark,
with two indistinguishable jets from the $W$ we lose this capability.
However having made a choice for defining $\theta_1$, the definition
of $\theta_2$ is correlated with that of $\theta_{12}$. The invariance
of the $\cN^{(qq)}$ distribution under this symmetry is manifest
by the dependence on the quadratic terms $\cos^2\theta_1$, $\cos^2\theta_2$
and $\cos\theta_1\cos\theta_2$ only.

As long as the photon
direction $\Vn$ is kept away from the ``dead cones'' of the final
charges, a simplified version of Eqs.~(\ref{NqqlA}) can be used in which
the velocities are set to 1:
\bminiG{NqqlAR}
\label{NqqAR}
\eqalign{
\cN^{(qq)}
&= \frac{\frac{1}{9}+  \cos^2\!\theta_1}{\sin^2\theta_1} +
\frac{\frac{1}{9}+  \cos^2\!\theta_2}{\sin^2\theta_2}
 + 2\chi(\omega)
\frac{(\cos\theta_1\cos\theta_2-\cos\theta_{12})
\cos\theta_1\cos\theta_2}{\sin^2\theta_1\sin^2\theta_2}
 ;  }  \\
\nonumber\\
\label{NqlAR}
\eqalign{
\cN^{(q\ell)}
&=  \frac{\frac{1}{9}+  \cos^2\theta_1}{\sin^2\theta_1} +
\frac{\sin^2 \theta_2 }{(1-\cos \theta_2)^2}
 + 2\chi(\omega)
\frac{(\cos\theta_1\cos\theta_2-\cos\theta_{12})
\cos\theta_1}{\sin^2\theta_1\>(1-\cos \theta_2)}\>. }
\emini
Expressing the relative angle $\theta_{12}$ between the final jets (the jet and
the lepton) in terms of photon angles with respect to the 1,2 directions,
we can write the final expression for the angular pattern of photon radiation
as
\bminiG{NqqlF}
\cN^{(qq)} &\propto&  \frac{\frac{1}{9}+\cos^2\theta_1}{\sin^2\theta_1} +
\frac{\frac{1}{9}+\cos^2\theta_2}{\sin^2\theta_2}
-2\chi(\omega)\>\cot\theta_1 \>\cot\theta_2 \cdot \cos\phi_{12}
\>,\\
\cN^{(q\ell)} &\propto& \frac{\frac{1}{9}+\cos^2\theta_1}{\sin^2\theta_1} +
\>\> \cot^2\frac{\theta_2}{2}\>\>
-2\chi(\omega) \>\cot\theta_1 \>{\rm{cotan}}\frac{\theta_2}{2}
\cdot \cos\phi_{12} \>,
\emini
where $\phi_{12}$ is the azimuthal angle between $\Vo$ and $\Vt$ projections
onto the plane {\em orthogonal}\/ to the photon direction, $\Vn$.

To study the $\theta_{12}$ dependence of the {\it total} photon yield
one has to evaluate the integrals over the gluon radiation angle
of the interference terms in (\ref{NqqlA}).
These integrals are finite at $v_1=v_2=1$ so we can use the
relativistic approximation (\ref{NqqlAR}) for this purpose.
A straightforward calculation
leads then to identical results for the quark-quark
and quark-lepton channels, namely
\beq
\eqalign{
&\int\frac{d\Omega}{4\pi}\>\> \frac{(\cos\theta_1\cos\theta_2-\cos\theta_{12})
\cos\theta_1\cos\theta_2}{\sin^2\theta_1\>\sin^2\theta_2} \cr
&=\int\frac{d\Omega}{4\pi}\>\> \frac{(\cos\theta_1\cos\theta_2-\cos\theta_{12})
\cos\theta_1}{\sin^2\theta_1\>(1-\cos \theta_2)} \>\>=
\ln\frac{\sin\theta_{12}}{2} \>+\>1 \>.
}\eeq
Taken together with the independent contributions, the total photon yield
then takes the form
given in Eqs. (\ref{cNall},\ref{cRilq}).

\newpage
\setcounter{equation}{0}
\section{Azimuthal angle integrations}
\def\atan{\mathop{\rm atan}}

In this Appendix we derive the expressions for the radiation
pattern in $t\bar{t}$ production when the gluon
 is integrated over its azimuthal
angle $\phi$  with respect to the direction of the $b$-quark ({\it i.e.}
the \lq\lq 1" direction).

We first note that the $\phi $ dependence enters when we substitute
\begin{equation}
\cos\theta_2 = \cos\theta_1 \cos\theta_{12} + \cos\phi
\sin\theta_1 \sin\theta_{12}
\end{equation}
into the result for $\cRc$ given in Eq. (\ref{coher}). Consider first
the azimuthal average of the interference term, $\lrang{\cJ }$.
We need the
basic integral  $\>( a \ge\abs{b})$
\beq\label{fullaz}
\int_0\pw{2\pi} \frac{d\phi}{2\pi}\> \frac1{a+b\cos\phi} =
\frac1{\sqrt{a\pw2-b\pw2}}\>.
\eeq
where
\beq
a\>=\> 1-v_2\cos\theta_1\cos\theta_{12}\>, \quad
b\>=\> -v_2\sin\theta_1\sin\theta_{12}\>.
\eeq
This gives
\beq
V \equiv\lrang{\frac{1}{1-v_2\cos\theta_2}} =
\left[A\pw2 + \sin^2\theta_1 (1-v_2\pw2)
\right]\pw{-\half}\>,
\eeq
where
\beq
A = \cos\theta_1 - v_2 \cos\theta_{12}\; ,
\eeq
which then leads to
\beq
\lrang{J} \> =\> \frac{v_1}{1- v_1 \cos\theta_1} \left[
V\cdot A - \cos\theta_1\right]\; .
\eeq
As a function of $\theta_1$
the first term in the bracket is a smooth step-function-like
distribution, falling from $+1$ at $\theta_1 = 0$, through $0$
at $\cos\theta_1 = v_2 \cos\theta_{12}$ ({\it i.e.} when
$\theta_1$ coincides with the direction of the other $b$-quark),
to $-1$ at $\theta_1 = \pi$, see also Ref. [\webber].

For the azimuthal average of the $\cRtwo$ contribution to $\cRc$
we can make use of the following:
\beq
V' \equiv \lrang{\frac{1}{(1-v_2\cos\theta_2)\pw2}} =
\left(v_2{\partial \over \partial{v_2}}+1\right) V =
(1-v_2\cos\theta_1\cos\theta_{12})\cdot V\pw3\> ,
\eeq
from which follows
\beq
\lrang{\cR_2} = 2V -1-(1-v_2\pw2)\,V'\>.
\eeq

Note that in the special case when $\theta_{12} = 0$ we obtain
\beq
\lrang{J} \> =\> -  \frac{v_1 v_2 \sin^2\theta_1}{(1- v_1 \cos\theta_1)
(1- v_2 \cos\theta_1)} \; ,
\eeq
and
\beq
\lrang{\cRc} \> =\>   \frac{(v_1- v_2)^2 \sin^2\theta_1}{(1- v_1
\cos\theta_1)^2
(1- v_2 \cos\theta_1)^2} \; .
\label{cRcav}
\eeq
Evidently $\lrang{\cRc}$ vanishes when $v_1 = v_2$, because
of {\it complete} destructive interference.

A similar analysis can be performed for the more symmetric case
of the azimuthal average around the {\it bisector} of the ``1"
and ``2" directions. If $\Theta$  is the angle
between the gluon and the bisector and $\Delta = \half \theta_{12}$
then a straightforward but tedious calculation gives for the independent
contributions
\beq
\lrang{\cR_i} = 2U_i-1-(1-v_i\pw2)U_i' \>,
\eeq
and for the coherent contribution
\beq
\lrang{\cRc}= \frac{2(1-v_1v_2\cos2\Delta)}
{v_1+v_2-2v_1v_2\cos\Theta\cos\Delta} \left[\>v_1U_1+v_2U_2\>\right]
- (1\!-\!v_1\pw2)U_1' - (1\!-\!v_2\pw2)U_2' \>,
\eeq
where
\bmini
U_i&\equiv&\lrang{\frac{1}{1-v_i\cos\Theta_i}} =
\left[(1-v_i\pw2)\sin\pw2\Delta + (\cos\Delta-v_i\cos\Theta)\pw2
\right]\pw{-\half}\>,\\
U_i' &\equiv&\lrang{\frac{1}{(1-v_i\cos\Theta_i)\pw2}} =
\left(v_i\partder{v_i}+1\right)U_i=
(1-v_i\cos\Theta\cos\Delta) U_i\pw3.
\emini

It can be further shown that for the equal velocities case ($v_1 = v_2 = v$)
\begin{eqnarray}
\lrang{\cRc}&=& \frac{2v\pw2\sin\pw2\Delta }
{1-v\cos\Theta\cos\Delta} \left[ 2 U\pw{-1} -\sin\pw2\Theta(1-v\pw2) \right]
U\pw3  \nonumber \\
&\  \propto& \ \Delta^2 \; ,\qquad \Delta \to 0 \; ,
\end{eqnarray}
while for $v_1 \neq v_2$ and $\theta_{12} = 0$ we again arrive at ({\it cf.}
Eq. (\ref{cRcav}))
\beq
\lrang{\cRc} \> =\>   \frac{(v_1- v_2)^2 \sin^2\Theta}{(1- v_1
\cos\Theta)^2
(1- v_2 \cos\Theta)^2} \; .
\eeq

\newpage
\noindent{{\Large\bf References}}
\begin{itemize}

\item[{[\tdkbis]}]
Yu.L. Dokshitzer, V.A. Khoze and S.I. Troyan, University of Lund
preprint LU-TP-92-10 (1992).

\item[{[\kos]}] V.A. Khoze,  L.H. Orr and W.J. Stirling,
Nucl.~Phys. {\bf B378} (1992) 413.

\item[{[\cdf]}] CDF collaboration: F. Abe {\it et al.},
Phys. Rev. {\bf D45} (1992) 3921.

\item[{[\kuhn]}]
J.H. K\"uhn, Acta Phys. Austr. Suppl. {\bf 24} (1982) 203.

\item[{[\bigi]}]
I.I. Bigi {\it et al.}, Phys. Lett. {\bf 181B} (1986) 157.

\item[{[\orrros]}]
L.H. Orr and J.L. Rosner, Phys. Lett. {\bf 246B} (1990) 221; {\bf 248} (1990)
474(E).

\item[{[\jikia]}]
G. Jikia, Phys. Lett. {\bf 257B} (1991) 196.

\item[{[\muta]}] T. Muta, R. Najima and S. Wakaizumi, Mod. Phys.
Lett. {\bf A1} (1986) 203.

\item[{[\altarelli]}] G. Altarelli {\it et al.} in {\it Physics at
LEP}, eds. J. Ellis and R. Peccei, CERN 86-02, vol.1, p.1 (1986). \\
E. Longo {\it et al.} in {\it ECFA Workshop on LEP200}, eds. A. B\"ohm
and W. Hoogland, CERN 87-08, vol.1, p.85 (1987).

\item[{[\sommerfeld]}] A. Sommerfeld, ``{\it Atombau und Spektrallinien}",
Bd. 2 (Vieweg, Braunschweig) 1939.

\item[{[\sakharov]}] A.D. Sakharov, JETP {\bf 18} (1948) 631.

\item[{[\gammaw]}] UA2 collaboration: J. Alitti {\it et al.},
Phys. Lett. {\bf 276B} (1992) 354.

\item[{[\wwpaper]}]
Yu.L. Dokshitzer, V.A. Khoze, L.H. Orr and W.J. Stirling,
in preparation.

\item[{[\book]}]
Yu.L. Dokshitzer, V.A. Khoze, A.H. Mueller, and S.I. Troyan,
{\it Basics of Perturbative QCD}, Editions Frontieres, 1991.

\item[{[\webber]}] G. Marchesini and B. R. Webber, Nucl. Phys. {\bf B330}
(1990) 261.

\item[{[\velt]}] M. Lemoine and M. Veltman, Nucl. Phys.
{\bf B164} (1980) 443.

\item[{[\ghadir]}] G. E. Abu Leil, University of Durham preprint
DTP/92/84 (1992).

\item[{[\lund]}] B. Andersson, G. Gustafson and T. Sj\"ostrand,
Phys. Lett. {\bf 94B} (1980) 211.

\item[{[\drag]}] Ya.I. Azimov {\it et al.},
Phys. Lett. {\bf 165B} (1985) 147.

\item[{[\oldqed]}] See for example: J. M. Jauch and F. Rohrlich,
{\it The theory of photons and electrons: the relativistic
quantum field theory of charged particles with
spin one-half}, (1976); A.I. Akhiezer and V.B. Berestetskii,
{\it Quantum electrodynamics}, Interscience (1965).

\item[{[\study]}] See for example the report of the Top Quark Physics
working group in the Proceedings of the Workshop on
$e\pw+e\pw-$ collisions at 500 GeV, ed. P.M. Zerwas, DESY 92-123A,
p. 255 (1992).

\item[{[\fujii]}] K. Fujii, KEK Preprint 92-6 (1992), and references therein.

\end{itemize}
\newpage
\noindent{{\Large\bf Figure Captions}}
\begin{itemize}
\item [{[1]}]
Soft gluon distribution in $\ee\to\tt\to\bb WW$ near $\tt$ threshold,
for gluons perpendicular to the $\bb$ plane, for $\chi$ as marked.
$dN/d\omega d\Omega$ is given by Eqs. (\ref{dN}) and (\ref{N}),
with $\theta_1=\theta_2=90\degree$.  $\t12$ is the angle between the $b$
and $\bbar$.  Here and in all subsequent figures, unless otherwise noted,
$v=0.9944$, which corresponds to $m_t=140$\ GeV for $m_b=5$\ GeV and
$m_W=80\ \GeV$.

\item [{[2]}]
Distribution of soft gluons radiated out of the $\bb$ plane, with energy and
angular integrations, for $\Gamma$ as marked.  The distribution shown is
given by
\begin{displaymath}
\int dN \equiv 2 \int_{5\ \GeV}^{10\ \GeV} d\omega
\int_{3\pi/8}^{5\pi/8}\sin\theta\; d\theta \int_{3\pi/8}^{5\pi/8} d\phi\;
(dN/d\omega d\Omega),
\end{displaymath}
where $\theta$ and $\phi$ are, respectively, the polar and azimuthal
angles with respect to the $b$ quark direction; the $\bb$ pair defines
the $xz$ plane.
The factor of $2$ accounts for
radiation on both sides of the plane.  $\Gamma=0.7\ \GeV$ (dotted line)
corresponds to a Standard Model top with $m_t=140\ \GeV$.

\item [{[3]}]
Angular ordering effects in the soft gluon distribution with azimuthal
integration:
\begin{displaymath}
\int dN \equiv \omega (dN/d\omega d\cos\theta) =
\int_{0}^{2\pi} d\phi\; \omega (dN/d\omega d\Omega).
\end{displaymath}
The angles between the $b$ and $\bbar$ are
(a) $\t12=5\degree$,
(b) $\t12=30\degree$,
(c) $\t12=90\degree$, and
(d) $\t12=120\degree$.
Solid lines: $\chi=0$; dotted lines: $\chi=0.5$; dashed lines: $\chi=1$.

\item [{[4]}]
Analog of Fig. 3 for the $\ww$ case:
angular ordering effects in the soft {\it photon} distribution in
$\gamma\gamma\to\ww\to l\pw+l\pw- \nu \bar\nu$, for
$\t12=30\degree$ and $150\degree$.
$\int dN \equiv \omega (dN/d\omega d\cos\theta)$ as in Fig. 3.
 Solid lines:  $\chi=0$;
dashed lines: $\chi=1$.
(The $\gamma\gamma\to\ww$ case is equivalent to the $\ee\to\ww$ case with no
initial state radiation.)

\item [{[5]}]
Soft gluon distribution (a) ``between'' and (b) ``outside'' the $\bb$ pair.
In (a),
\begin{displaymath}
\int dN \equiv  \int_{5\ \GeV}^{10\ \GeV} d\omega
\int_{0}^{2\pi} d\phi \int_{0}^{\t12}\sin\theta d\theta
(dN/d\omega d\Omega),
\end{displaymath}
and in (b),
\begin{displaymath}
\int dN \equiv  \int_{5\ \GeV}^{10\ \GeV} d\omega
\int_{0}^{2\pi} d\phi \int_{\t12}^{\pi}\sin\theta d\theta
(dN/d\omega d\Omega).
\end{displaymath}
Solid lines:  $\Gamma=0, \infty$ (as marked);
dotted line:  $\Gamma=0.7\ \GeV$;
dashed line:  $\Gamma=3\ \GeV$;
dot-dashed line:  $\Gamma=5\ \GeV$.

\item [{[6]}]
Net emission probability for gluons with energies from 5 to 10 GeV:
\begin{displaymath}
\int dN \equiv  \int_{5\ \GeV}^{10\ \GeV} d\omega
\int_{0}^{2\pi} d\phi \int_{0}^{\pi}\sin\theta d\theta
\; (dN/d\omega d\Omega).
\end{displaymath}

\end{itemize}

\end{document}